\documentstyle[12pt]{article}

\def\eq#1{Eq.~(\ref{#1})}
\def\eqs#1#2{Eqs.~(\ref{#1}) and (\ref{#2})}

\def\eqst#1#2{Eqs.~(\ref{#1})--(\ref{#2})}

\def\rfrac#1#2{\left(\frac{#1}{#2}\right)}

\def\text#1{\hbox{\rm#1}}

\newcommand{\sub}[1]{_{\mbox{\rm\scriptsize#1}}}

\def\pr{^\prime}

\def\ee{\end{equation}}
\def\be{\begin{equation}}
\def\eea{\end{eqnarray}}
\def\bea{\begin{eqnarray}}
\def\eqa{\!\!\!&=&\!\!\!}

\def\simeqa{\!\!\!&\simeq &\!\!\!}

\def\km{\,\text{km}}

\def\Mpc{\,\text{Mpc}}

\def\mone{^{-1}}
\def\mtwo{^{-2}}
\def\mthree{^{-3}}
\def\mfour{^{-4}}

\def\mhalf{^{-1/2}}
\def\mthreehalf{^{-3/2}}
\def\half{^{1/2}}
\def\threehalf{^{3/2}}

\def\twothird{^{2/3}}

\def\mthird{^{-1/3}}

\def\pr{^\prime}

\def\msun{M_\odot}

\def\d{\text d}

\def\s{\text s}

\def\bfa#1{{\bf #1}}

\def\del{\mbox{\boldmath$\nabla$} }

\def\pa{\partial}
\def\pd#1#2{\frac{\pa #1}{\pa #2}}

\def\lsim{\mathrel{\rlap{\lower4pt\hbox{\hskip1pt$\sim$}}
    \raise1pt\hbox{$<$}}}
\def\gsim{\mathrel{\rlap{\lower4pt\hbox{\hskip1pt$\sim$}}
    \raise1pt\hbox{$>$}}}

\def\calp{{\cal P}}

\begin{document}
  \begin{titlepage}
  \begin{flushright}
SUSSEX-AST 92/8-1; LANC-TH 8-92\\
(August 1992)\\
  \end{flushright}
  \begin{center}
{\Large
{\bf The Spectral Slope in the Cold Dark Matter Cosmogony\\}}
\vspace{.3in}
{\large Andrew R.~Liddle$^{\dagger}$ and David H.~Lyth$^*$\\}
\vspace{.4 cm}
{\em  $^{\dagger}$Astronomy Centre, \\ Division of Physics and Astronomy, \\
University of Sussex, \\ Brighton BN1 9QH.~~~U.~K.}\\
\vspace{.4 cm}
{\em  $^*$School of Physics and Materials,\\ Lancaster University,\\
Lancaster LA1 4YB.~~~U.~K.}\\
\end{center}
\baselineskip=24pt
\begin{abstract}
\noindent
In a recent paper, we suggested that the density fluctuation spectra arising
from power-law (or extended) inflation, which are tilted with respect to the
Harrison--Zel'dovich spectrum, may provide an explanation for the excess large
scale clustering seen in galaxy surveys such as the APM survey. In the light
of the new results from COBE, we examine in detail here cold dark matter
cosmogonies based on inflationary models predicting power-law spectra. Along
with power-law and extended inflation, this class includes natural inflation.
The latter is of interest because, unlike the first two, it produces a
power-law spectrum without significant gravitational wave production. We
examine a range of phenomena, including large angle microwave background
fluctuations, clustering in the galaxy distribution, bulk peculiar velocity
flows, the formation of high redshift quasars and the epoch of structure
formation.

Of the three models, only natural inflation seems capable of explaining the
large scale clustering of optical galaxies. Such a model, though at best
marginal even at present, has some advantages over standard CDM and on most
grounds appears to perform at least as well. Power-law inflation's primary
interest may ultimately only be in permitting a larger bias parameter than
standard CDM; it appears unable to explain excess clustering. Most models of
extended inflation are ruled out at a high confidence level.
\end{abstract}

\vspace{0.4cm}
E-mail addresses: arl @ uk.ac.sussex.starlink~~;~~d.lyth @ uk.ac.lancaster
%%%%%%%%%%%%%%%%%%%%%%%%%%%%%%%%%%%%%%%%%%%%%%%%%%%%%%%%%%%%%%%%%%%%%%
\end{titlepage}
%%%%%%%%%%%%%%%%%%%%%%%%%%%%%%%%%%%%%%%%%%%%%%%%%%%%%%%%%%%%%%%%%%%%%%
%%%%%%%%%%%%%%%%%%%%%%%%%%%%%%%%%%%%%%%%%%%%%%%%%%%%%%%%%%%%%%%%%%%%%%

\section{Introduction}
\label{INTRO}
\setcounter{equation}{0}
\def\theequation{\thesection.\arabic{equation}}

Although recently it has been running into trouble from several sources, the
standard Cold Dark Matter (CDM) model (White {\it et al} 1987; Frenk {\it et
al} 1988; Efstathiou 1990) has shown some considerable success in explaining
the clustering of galaxies we see around us. This model is based upon the idea
that there exists a primordial spectrum of density fluctuations which takes on
a particular form known as the Harrison--Zel'dovich or flat spectrum. Most
frequently, this choice is motivated as being a prediction of the inflationary
universe model. Nevertheless, it has long been known, at least within the
inflation community, that the predicted spectrum is only approximately flat,
and that within any given model there are readily calculable deviations from
scale invariance. An example is provided by the standard `chaotic' inflation
scenario (Linde 1983, 1987, 1990), wherein a logarithmic correction (see {\it
eg} Salopek, Bond \& Bardeen 1989; Mukhanov, Feldman \& Brandenberger 1992;
Schaefer \& Shafi 1992) adds some power at large scales, giving a density
contrast at horizon crossing of 5--10\% greater at $1000h^{-1}$ Mpc than at
$10h^{-1}$ Mpc. Hitherto, when our knowledge of the primeval spectrum was
restricted to studies of local clustering up to scales of perhaps $100 h^{-1}$
Mpc, these corrections where rightly regarded as insignificant. The recent
results (Smoot {\it et al} 1992) from the Cosmic Background Explorer satellite
(COBE) have changed this picture, by providing for the first time good
estimates of the amplitude of the spectrum on scales of upwards of $1000
h^{-1}$ Mpc. The time has come for these correction terms to be taken
seriously, and possibly the real test of inflation lies not in how flat the
primeval spectrum is, but in the type and size of deviations from it.

A certain class of inflationary models predict an even more dramatic change
from the Harrison--Zel'dovich case, wherein the primeval spectrum generated
exhibits power-law deviations from flatness. There are several such
inflationary scenarios. The simplest is power-law inflation (Abbott \& Wise
1984b; Lucchin \& Matarrese 1985), which can be realised via a scalar field
evolving in a potential of exponential form and which leads to an expansion of
the universe which is not the conventional nearly exponential growth, but
rather has the scale factor growing as a rapid power law in time. A second
option is provided by the so-called extended inflation scenarios (La \&
Steinhardt 1989; Kolb, Salopek \& Turner 1990; Kolb 1991), wherein
modifications to Einsteinian gravity allow inflation to proceed via a
first-order phase transition. Yet another option is provided by `natural
inflation' (Freese, Frieman \& Olinto 1990; Adams {\it et al} 1992), a
specific case of the more general situation of a scalar field evolving near
the top of an inverted harmonic oscillator potential which also gives a
power-law spectrum.

While power-law primeval spectra are often written down, the consequences of
such a spectrum in the complete CDM cosmogony have not been fully
investigated. The first detailed discussion of such models was made by
Vittorio, Matarrese and Lucchin (1988), where they also allowed the density
parameter $\Omega$ to be less than one. Subsequent investigations by these
authors and collaborators (Tormen, Lucchin \& Matarrese 1992; Tormen {\it et
al} 1992) have concentrated primarily on analysis of bulk velocity flows.
Interesting constraints in the $n$--$\Omega$ plane were also obtained by Suto
and Fujita (1990) and by Suto, Gouda and Sugiyama (1990) based on predictions
for the cosmic Mach number combined with the then current microwave anisotropy
limits. A pre--COBE analysis much less specific than that given here has been
given by Kashlinsky (1992). In a recent paper with Will Sutherland (Liddle,
Lyth \& Sutherland 1992, henceforth LLS), we examined the consequences of such
spectra for galaxy clustering. The motivation lay in recent observations such
as the APM survey (Maddox {\it et al} 1990, 1991), which have shown that on
large scales there appears to be stronger clustering than the standard CDM
model can provide. These power-law spectra do indeed possess extra power on
large scales. We found that for a certain range of values of the slope, one
could reproduce the clustering seen in the APM survey, while withstanding the
then current microwave background bounds.

Given this success, the emergence of the COBE data makes this an excellent
time to examine this scenario in more detail, especially as in a certain sense
the COBE data too show a rather larger amplitude of fluctuations than one
would expect in the standard biassed CDM model. We have also taken this
opportunity to analyse several other aspects of the scenario, though we have
not gone so far as to carry out $N$-body simulations. Instead, we have
restricted ourselves to calculations which can be done analytically or
semi-analytically.

Recently there have been several papers analysing various aspects of power-law
spectra (Salopek 1992; Davis {\it et al} 1992; Liddle \& Lyth 1992; Lucchin,
Matarrese \& Mollerach 1992; Adams {\it et al} 1992; Lidsey \& Coles 1992),
which have however tended to deal with specific issues relating to particular
models. With the exception of Adams {\it et al}, these papers have also failed
to come up with detailed constraints on the scenario. Our aim here is to
provide a more unified treatment, covering all the inflationary possibilities
for generating power-law spectra, with the ultimate aim of constraining the
slope of the spectrum and the bias parameter for the different models. One has
already seen limits on the slope quoted by COBE (Smoot {\it et al} 1992) of $n
= 1.1 \pm 0.5$, these numbers being simply derived from the variation of
microwave anisotropies with scale between around $10^0$ and the quadrupole,
and thus being independent of the specific choice of dark matter. [It has been
suggested (D.~T.~Wilkinson, conference talk, Oxford, June 1992) that a more
detailed error analysis (recognising that the error bars on the data are not
independent) will {\em increase} the size of the errors on $n$.] It is natural
for one to expect that if one makes the additional requirement, within a
specific cosmogony, that the appropriate galaxy clustering is obtained at
scales a factor of 100 smaller than those COBE samples, then the limits on the
spectral slope become yet tighter. On the other hand, one should bear in mind
that the COBE error bars on $n$ are presumably, like the rest of the data,
$1$-sigma error bars and so a stronger level of exclusion would allow further
deviations from scale invariance.

Our general aim is to constrain the spectral index and bias parameter in the
cold dark matter cosmogony by finding limits on the spectral amplitude at
different length scales. Of most interest for this purpose will be the
amplitude on large scales ($>1000h^{-1}$ Mpc) from the COBE experiment, and on
intermediate scales using the velocity data (which see the mass spectrum
itself rather than the galaxy spectrum). In particular we shall use results
from the QDOT survey (Saunders {\it et al} 1991; Saunders, Rowan-Robinson \&
Lawrence 1992) which fix the amplitude at around $20 h^{-1}$ Mpc. It is
important to note that both these scales are expected to be well into the
linear regime and thus linear calculational techniques can be used. To aid
comparison with other work, we shall normally give the inferred normalisation
in terms of the optical galaxy bias parameter $b$ defined later, which is
assumed to be the reciprocal of the mass dispersion in $8 h^{-1}$ Mpc spheres
using the CDM spectrum. Note that this use should not be taken to imply that
the results have anything to do with optical galaxies or scales which are
close to entering the nonlinear regime.

Having obtained limits in this way on the mass spectrum, reference to optical
galaxies provides limits on the bias parameter, enabling us to present our
constraints in the form of allowed regions in the $n$--$b$ plane. In order to
clarify why particular choices of $n$ appear in these sections, let us presage
here the ultimate result that $n$ is constrained in any inflation model to be
no less than $0.7$, at very high confidence. With the allowed values in mind,
we then go on to investigate whether any further constraints can be imposed by
an examination of aspects of the formation of nonlinear structure, wherein
however the calculations are technically more tricky and comparison with
observation much more ambiguous.

The layout of this paper is thus as follows, and generally works from large to
small scales. Firstly, section \ref{INFL} discusses the inflationary models
giving rise to power-law spectra. In section \ref{SPECTRUM}, we discuss our
processed CDM spectrum and its normalisation. Our aim here is to be as
explicit as possible in our definitions and in where the uncertainties lie.
Section \ref{MWB} discusses large angle microwave background anisotropies, a
discussion dominated of course by the observations of the COBE satellite.
These calculations refer to the largest scales inside our horizon. In section
\ref{CLUSTER}, we review the results of LLS concerning the application of
power-law spectra to the excess large-scale galaxy clustering. In this section
we also discuss the bulk flows, which serve as another indicator of the power
spectrum amplitude on these scales. Section \ref{NONLIN} deals with the issue
of shorter scales, corresponding to typical galaxy and cluster masses. There
we investigate a range of topics concerning the entry into the nonlinear
regime In this region of the spectrum, our power-law spectra offer less power
than standard CDM, and hence one expects the structure formation constraint to
be more stringent here. Finally, we conclude and discuss our results.

The universe is taken to have critical density, with present Hubble parameter
$H_0=100 h \km\,\s\mone\Mpc$ where $h=0.5$. For ease of comparison with
other work we often take $h\mone\Mpc$ for the distance unit. The
units are taken to be such that $c=1$ and (in section \ref{INFL}) $\hbar=1$,
and the Planck scale is defined by $m_{Pl}=G\mhalf$.

\section{Inflationary Models giving Power-Law Spectra}
\label{INFL}
\setcounter{equation}{0}
\def\theequation{\thesection.\arabic{equation}}

As discussed in the introduction, there are at least three different
inflationary models giving rise to power-law spectra. While there are
certainly relations between these as we shall see, each has its own features
and we shall discuss each in turn.

Let us first briefly sketch the derivation of the spectra, following Liddle
and Lyth (1992). A scalar field $\phi$ moving in a potential $V$ is usually
taken to satisfy slow-roll conditions ($' \equiv {\rm d}/{\rm d}\phi$)
\begin{eqnarray}
\dot{\phi} = -\frac{1}{3H} V'\\
\epsilon \equiv \frac{m_{Pl}^2}{16\pi} \left( \frac{V'}{V} \right)^2 \ll 1 \\
\eta \equiv \frac{m_{Pl}^2}{8\pi} \frac{V''}{V} \ll 1
\end{eqnarray}
$H \equiv \dot{a}/a$ is the Hubble parameter during inflation, where $a$ is
the scale factor normalised to $a=1$ at present. Denoting the epoch at the end
of inflation by a subscript $2$, and the epoch $k=aH$ when a scale $k$ leaves
the horizon with a subscript $k$, the number of Hubble times $N(t_k)$ between
a scale $k$ leaving the horizon and the end of inflation is
\begin{equation}
N(t_k) = \frac{8\pi}{m_{Pl}^2} \int_{\phi_k}^{\phi_2} \frac{V}{V'} {\rm d}\phi
	= 60 + \ln \frac{V_k^{1/4}}{10^{16} {\rm GeV}} + 2 \ln
	\frac{V_k^{1/4}}{V_2^{1/4}} - \ln (k/H_0)
\end{equation}

In the slow-roll approximation, the spectrum of density fluctuations is given
by
\begin{equation}
\delta_H^2(k) = \frac{32}{75} \frac{V_k}{m_{Pl}^4} \epsilon_k^{-1}
\end{equation}
where $\delta_H$ is roughly the density contrast at horizon entry and is
defined precisely in the following section. The effective spectral index is
given by
\begin{equation}
1-n \equiv -\frac{{\rm d} \ln[\delta_H^2(k)]}{{\rm d} \ln k} =
\left( 6 \epsilon_k - 2 \eta_k \right)
\end{equation}
We see immediately that in order to have significant deviations from the flat
$n=1$ spectrum one must come close to violating at least one of the slow-roll
conditions. Corrections to these slow-roll results have been calculated
exactly for power-law inflation (Lyth \& Stewart 1992a), and to first order in
the departure from slow-roll for the general case (Lyth \& Stewart 1992b). As
discussed in our earlier paper (Liddle \& Lyth 1992) they are not important in
practice.

The models giving rise to power-law spectra are as follows.

\begin{enumerate}
\item {\em Power-Law Inflation.}

Power-law inflation $a \propto t^p$ arises (Abbott \& Wise 1984b; Lucchin \&
Matarrese 1985; Barrow 1987; Liddle 1989) when a scalar field $\phi$ evolves
down an exponential potential of the form
\begin{equation}
V(\phi) = V_0 \exp \left( \sqrt{\frac{16\pi}{p\, m_{Pl}^2}} \, \phi \right)
\end{equation}
It is of particular interest to researchers in inflation because although the
scalar field comes close to violating the slow-roll conditions, exact analytic
solutions exist both for the dynamics of inflation and for the density
perturbations generated. The exact formula for the spectral index is (Lyth \&
Stewart 1992a)
\begin{equation}
n = 1 - 2/(p-1)
\end{equation}
In the limit of large $p$ the spectrum tends rapidly towards the flat $n=1$
spectrum.

\item {\em Extended Inflation.}

Extended inflation (La \& Steinhardt 1989; Kolb 1991) looks very different
from power-law inflation, but is in fact related during inflation by a
conformal transformation which allows the usual power-law inflation machinery
to be utilised in its study. It is based on modifications to the gravitational
sector of the theory, which allow a first-order inflationary phase transition
to complete satisfactorily. The original model (La \& Steinhardt 1989) was
based on a Brans-Dicke theory with parameter $\omega$, and although this
proved insufficient to allow present day tests of general relativity to be
satisfied, it has remained the paradigm around which more complicated working
models are based. The spectral slope in such a model is given by (Kolb,
Salopek \& Turner 1990; Guth \& Jain 1992; Lyth \& Stewart 1992a)
\begin{equation}
n = \frac{2\omega - 9}{2\omega - 1}
\end{equation}
with extended inflation being conformally equivalent to power-law inflation
with $2p = \omega + 3/2$.

A crucial difference between power-law and extended inflation is that extended
inflation suffers an additional constraint, as one must avoid the large
bubbles generated as inflation ends from being so profuse as to unacceptably
distort the microwave background. This constrains $\omega$ as a function of
the inflaton energy scale $M$ as (Liddle \& Wands 1991; Liddle \& Lyth 1992)
\begin{equation}
\omega < 20 + 0.7 \log_{10} \left( M/m_{Pl} \right)
\end{equation}
Bounding $M$ using the microwave background limits on the fluctuation amplitude
gives $\omega \lsim 17$, corresponding to $n \lsim 0.75$. We have discussed the
significance of this bound in an earlier paper (Liddle \& Lyth 1992); the
prognosis for the extended inflation model is not good, as we shall recap in
this paper.

\item {\em Natural Inflation.}

The natural inflation model (Freese, Frieman \& Olinto 1990; Adams {\it et al}
1992) is based on a pseudo-Nambu-Goldstone boson evolving in a potential
\begin{equation}
V(\phi) = \Lambda^4 \left( 1 \pm \cos (\phi/f) \right)
\end{equation}
where $\Lambda$ and $f$ are mass scales. Evolution near the top of this
potential (as is required for sufficient inflation with parameter choices
significantly tilting the spectrum), gives rise to a power-law spectrum. In
fact, this is a realisation of the more general case of a scalar field
evolving near the top of an inverted harmonic oscillator potential
\begin{equation}
V(\phi) = V_0 - \frac{1}{2} m^2 \phi^2
\end{equation}
which gives a spectrum of slope
\begin{equation}
n = 1 - \frac{m^2 m_{Pl}^2}{4\pi V_0}
\end{equation}
For natural inflation one has $n = 1-m_{Pl}^2/8\pi f^2$.

That natural inflation gives the same spectrum as power-law inflation may seem
surprising in the light of claims that one can reconstruct inflationary
potentials from a given spectrum (Hodges \& Blumenthal 1990). In fact, these
two models can be regarded as different regimes of an all-encompassing
potential (which one can calculate in the manner of Hodges and Blumenthal)
which is essentially $1/\cosh^2 \phi$ with various factors thrown in. In the
$\phi \simeq 0$ region we have the inverted harmonic oscillator, while at
large $\phi$ we have the exponential region. In a sense, this is the unique
potential from which all inflationary models giving power-law spectra arise.
With this potential, the scalar field can roll from the top down to the
exponential region, while in a slow-roll approximation generating an exact
power-law spectrum. However, one must also note that as discussed above, the
fact that the spectrum has been tilted implies that the slow-roll
approximations are at best only just satisfied, so there will be corrections
to the slow-roll spectrum in all regions of this potential. In the exponential
region these corrections are known to affect only the amplitude and not the
slope; one hopes for the same near the top of the potential.
\end{enumerate}

Each of the above models can produce a power-law spectrum of the appropriate
amplitude. However, as discussed in many recent papers (Krauss and White 1992;
Salopek 1992; Davis {\it et al} 1992; Liddle \& Lyth 1992; Lucchin, Matarrese
\& Mollerach 1992; Souradeep \& Sahni 1992), one must pay attention to the
predicted amplitude of gravitational waves (this was noted for power-law
inflation long ago by Fabbri, Lucchin and Matarrese (1986)). In regimes where
the slow-roll approximation is invalid or close to violation, one expects that
these can become large, and the above work has confirmed that for power-law
inflation (and by inference for extended inflation), the microwave
contributions from gravitational waves may not only be large, but can dominate
those from the scalar density fluctuations. To a good enough approximation for
our purposes (see Lucchin, Matarrese \& Mollerach (1992) for more precise
results), the ratio $R$ of the tensor to the scalar contribution to the
squared microwave multipoles is independent of the multipole and given by
(Liddle \& Lyth 1992)
\begin{equation}
R \simeq 12.4 \frac{m_{Pl}^2}{16\pi} \left( \frac{V'}{V} \right)^2
\end{equation}
where the potential and its derivative are to be evaluated at the point
corresponding to a given scale passing out of the horizon. In general $R$ is
scale-dependent, but for power-law and extended inflation one has the
scale-independent ratio
\begin{equation}
R \simeq 12.4/p
\end{equation}
and thus for values of $p$ less than around $12$ ($n \lsim 0.82$) the tensor
modes dominate the observed microwave background anisotropies.

On the other hand, for natural inflation the ratio $R$ is extremely small as
the relevant scales leave the horizon, and hence there are no gravitational
wave corrections of note in that case. Thus natural inflation (and similar
models) provide an inflationary mechanism for generating power-law spectra
without the large gravitational wave accompaniment. We shall return to the
role of gravitational waves in section \ref{MWB} for their microwave
background implications, but for now we shall turn to the spectrum of density
fluctuations.

\section{The Spectrum}
\label{SPECTRUM}
\setcounter{equation}{0}
\def\theequation{\thesection.\arabic{equation}}

\subsection{Definitions}

The matter density contrast is $\delta(\bfa x,t) \equiv (\rho(\bfa
x)-\bar{\rho})/\bar{\rho}$, where $\rho(\bfa x)$ is the matter density,
$\bar\rho$ is its spatial average, $t$ is time and $\bfa x\equiv(x^1,x^2,x^3)$
are comoving Cartesian coordinates. There is some freedom in the definition of
$t$ and $\bfa x$, but this `gauge freedom' is not significant during the
matter dominated era which concerns us (Lyth \& Stewart 1990). As long as
they are small, $\delta$ and related quantities evolve according to the linear
equations of cosmological perturbation theory, with each Fourier mode evolving
independently. At each comoving point during the matter dominated era,
$\delta$ is proportional to the scale factor $a$, or equivalently to
$(aH)^{-2}$.

According to the inflationary models described in the last section the
primeval density contrast is a Gaussian random field at very early times, and
therefore as long as linear evolution holds. As such its stochastic properties
are completely determined by its spectrum. Different definitions of the
spectrum exist in the literature. Letting $\delta_{\bf k}$ denote the Fourier
coefficient of the density contrast in a box of coordinate volume $V$, with
the normalisation $\delta_{\bf k} = V\mone\int e^{-ik.x} \delta(x) d^3x$, we
work with the spectrum $\calp$ defined by
\begin{equation}
\calp = V(k^3/2\pi^2) \langle|\delta_\bfa k|^2\rangle
\end{equation}
where the bracket denotes the average over a small region of $k$-space. With
this definition the dispersion $\sigma$ of the density contrast is given by
\be
\label{DISP}
\sigma^2 = \int^\infty_0 \calp(k) \frac{dk}{k}
\ee
In the literature, our $\calp$ is denoted variously by $\calp_\rho$ (Salopek,
Bond and Bardeen 1989), $P_\rho$ (Lyth \& Stewart 1990), $\Delta^2$ (Kolb \&
Turner 1990), $\delta\rho_k/\rho$ (Linde 1990) and $\d\sigma^2_\rho/\d\ln k$
(Bond \& Efstathiou 1991). The quantity usually denoted by $P$ in the
literature is equal to $Nk\mthree \calp$, where $N$ is an author-dependent
normalisation factor which is often left undefined.

The standard assumption for the $k$-dependence of the primordial spectrum is
$P\propto k$, corresponding to $\calp\propto k^4$. This is variously referred
to as a Harrison--Zel'dovich spectrum, a flat spectrum, or a scale-free
spectrum, and it is generated by the vacuum fluctuation of the inflaton field
if the slow-roll conditions are well satisfied. Here we are exploring a
power-law primordial spectrum $P\propto k^n$, corresponding to $\calp\propto
k^{3+n}$, with $n<1$.

During the matter dominated era, the spectrum may be written
\begin{equation}
\calp(k) =\rfrac{k}{aH}^4 T^2(k) \delta_H^2(k)
\end{equation}
Up to a gauge dependent numerical factor (Lyth \& Stewart 1990), $\delta_H^2$
is the primeval spectrum at the epoch $k=aH$ when the scale $k$ leaves the
horizon. (In the literature $\delta_H(k)$ is often written as $\delta\rho/\rho
\! \! \mid_{hor}$.) Its scale dependence is $\delta_H^2\propto k^{n-1}$, which
is why the standard choice $n=1$ is referred to as a flat or scale--invariant
spectrum.

The quantity $T(k)$ is known as the transfer function. It measures distortions
to the primeval spectrum generated once a given scale comes within the
horizon. On those very large scales which enter the horizon long after matter
domination ($k\mone\gg 100\Mpc$) it is equal to 1. To calculate it on smaller
scales one has to know the matter content of the universe (see Efstathiou
(1990) for a discussion covering various physical situations). For hot dark
matter, free streaming erases perturbations on scales up to around the horizon
scale at matter-radiation equality, but for cold dark matter the predominant
effect is simply that the growth of modes which enter the horizon during
radiation domination is suppressed.

Taking the dark matter to be cold, the standard picture of the early universe
determines a practically unique transfer function once the baryon density
$\Omega_B$ is fixed. There are however considerable discrepancies between the
parametrisations which appear in the literature. For $\Omega_B=0$  there are
two widely quoted parametrisations, one due to Bond and Efstathiou (1984) and
the other to Bardeen {\it et al} (1986). We use the former, as given by
Efstathiou (1990)
\begin{equation}
T(k) = \left[ 1+\left( ak+\left(bk\right)^{3/2} + \left(ck\right)^2
	\right)^{\nu}\right]^{-1/\nu} \label{tran}
\end{equation}
where
\begin{equation}
a = 6.4 \left(\Omega h^2 \right)^{-1} {\rm Mpc} \; \; ; \; \; b=3\left(\Omega
	h^2\right)^{-1} {\rm Mpc} \; \; ; \; \; c=1.7\left(\Omega h^2
	\right)^{-1} {\rm Mpc} \; \; ; \; \; \nu=1.13
\end{equation}
In figure 1 this parametrisation is compared with that of Bardeen {\it et al}
(1986), and with two others. At scales around $k = 2\pi/10h^{-1}$ Mpc$^{-1}$,
the discrepancies can be of order 10\%. In particular, the parametrisation of
Davis {\it et al} (1985), which has the fewest fitting parameters, appears
rather less accurate than the others.

Adams {\it et al} (1992) claim that Bardeen {\it et al}'s parametrisation is
extremely accurate for $\Omega_B=0$, and that increasing $\Omega_B$ to the
value $\Omega_B\simeq.06$ favoured by nucleosynthesis decreases it by about
15\% at $k\simeq 2\pi/10h\mone\Mpc$. This would bring it 5\% below the
parametrisation that we have adopted, forcing the normalisation up by roughly
that amount.

\subsection{The filtered density contrast}

The density contrast $\delta(\bfa x)$ will evolve linearly as long as its
dispersion satisfies $\sigma\ll 1$, except in those rare regions where it
becomes $\gsim 1$ and gravitational collapse takes place. Following the usual
practice, we assume when necessary that the linear evolution is at least
roughly valid right up to the epoch $\sigma\simeq 1$. Soon after that epoch, a
large fraction of the matter collapses into gravitationally bound objects, and
linear evolution becomes completely  invalid. The epoch of non-linearity,
defined as the epoch when $\sigma =1$, corresponds to a redshift $z\sub{nl}$
given by $(1+z\sub{nl})=1/\sigma_0$, where $\sigma_0 $ is the \it linearly
evolved \rm quantity evaluated at the present time.

A `filtered' density contrast $\delta(R_f,\bfa x)$, which has a smaller
dispersion $\sigma(R_f)$, can be constructed by cutting off
the Fourier expansion of $\delta(\bfa x)$
above some minimum wavenumber $\simeq 1/R_f$, or
equivalently by smearing it over a region with size $\simeq R_f$. The
filtered quantity will evolve linearly until the later epoch $\sigma
(R_f)=1$.

A precise definition of the filtered density contrast is made by means of a
`window function' $W(r)$, which is equal to 1 at $r=0$ and which falls off
rapidly beyond some radius $R_f$ (Peebles 1980, Kolb \& Turner 1990).
The filtered density contrast is
\be \delta(R_f,\bfa x)=\int W(R_f,|\bfa x\pr-\bfa x|)
	\delta(\bfa x\pr) d^3 x\pr \ee
and its spectrum is
\be \calp(R_f,k)= \left[\widetilde W(R_f,k)/V_f\right]^2 \calp(k) \ee
where
\be \widetilde W(R_f,k)=\int e^{-i\bfa k\cdot\bfa x} W(R_f,r) d^3x \ee
and
\be V_f=\int W(R_f,r) d^3x \ee
The filtered dispersion is
\begin{equation}
\sigma^2(R_f)
= \int_0^{\infty} \left[\widetilde W(R_f,k)/V_f\right]^2
\calp(k) \frac{{\rm d}k}{k}
\end{equation}

The quantity $V_f$ is the volume `enclosed' by the filter. It is convenient to
define the associated mass $M=\rho_0 V_f$, where $\rho_0$ is the mean
comoving density which in an $h = 0.5$ universe is given by $\rho_0 =
3H_0^2/8\pi G = 6.94 \times 10^{10} \msun$ Mpc$^{-3}$ ($\msun$ being the solar
mass). One normally uses $M$ instead of $R_f$ to specify the scale, writing
$\delta(M,\bfa x)$ and $\sigma(M)$.

The two popular choices are the Gaussian filter
\bea W(R_f,r) \eqa \exp(-r^2/2 R_f^2) \label{gone} \\
V_f \eqa (2\pi)\threehalf R_f^3 \\
\widetilde W(R_f,k)/V_f \eqa \exp(-k R_f) \\
M \eqa 4.36\times 10^{12} h^2 (R_f/1\Mpc)^3 \msun
\label{gtwo} \eea
and the top hat filter which smears uniformly over a sphere of radius $R_f$
\bea W(R_f,r)\eqa\theta(r-R_f) \\
V_f \eqa 4\pi R_f^3/3\\
\widetilde W(R_f,k)/V_f \eqa 3 \left( \frac{\sin(k R_f)}{(k R_f)^3}
- -\frac{\cos(k R_f)}{(k R_f)^2} \right) \\
M \eqa 1.16\times 10^{12} h^2 (R_f/1\Mpc)^3 \msun
\eea
The Gaussian filter is the most convenient for the theoretical calculations
described in Section \ref{NONLIN}, but the top hat filter is widely used to as
a means of presenting data as in Section \ref{CLUSTER}.

\subsection{The conventional normalisation}

The final ingredient required to fix the CDM spectrum is the overall
normalisation. In the past this has usually been done by comparing the
linearly evolved theory with observation, for the galaxy correlation function
on a scale of order $10 h^{-1}$ Mpc. Two standard prescriptions exist
(Efstathiou 1990). The first requires that the dispersion $\sigma_g(r)$ of
galaxy counts in spheres of radius $r$ has the observed value unity at
$r=8h\mone\Mpc$ (Davis \& Peebles 1983). The second requires that $J_3(r)$,
the integral of the second moment of the galaxy correlation function up to
distance $r$, has the observed value (Davis \& Peebles 1983) 270 $h^{-3}$
Mpc$^{-3}$ at $r=10h^{-1}$ Mpc.

These normalisation schemes both refer to the statistics of the clustering of
luminous galaxies, not to the clustering of mass as described by the spectrum
$\calp$. It is an important ingredient of the CDM cosmogony that these are not
necessarily taken to be the same --- so-called biassed galaxy formation
(Bardeen {\it et al} 1986; Efstathiou 1990). To be specific the
mass density correlation function $\xi$, related to the spectrum by
\begin{equation}
\xi(r) = \int_0^{\infty} \calp (k) \frac{\sin kr}{kr} \frac{{\rm d}k}{k}
\end{equation}
is assumed to be related to the correlation function $\xi_{gg}$ of luminous
galaxies by
\be \xi_{gg}(r) \simeq b^2 \xi(r) \label{bias} \ee
with a
bias parameter $b$ which is roughly independent of the scale $r$.
Including the bias parameter, the normalisations implied by the two
schemes are
\bea \sigma_g^2(r)/b^2
\eqa 9 \int_0^{\infty} \calp_0(k)
	\left[ \frac{\sin kr}{(kr)^3} - \frac{\cos kr}{(kr)^2}
	\right]^2 \frac{{\rm d}k}{k}\nonumber\\
\eqa (1/b)^2 \text{\hspace*{1cm} at\ }r=8h\mone\Mpc
\label{sig} \eea
and
\bea
2r\mthree J_3(r)/b^2 \eqa \int_0^{\infty} 2 \calp_0(k)
\rfrac{1}{k r}^2
\left[\frac{\sin kr}{kr}-
	\cos kr \right] \frac{ {\rm d} k}{k}\nonumber\\
\eqa (.73/b)^2 \text{\hspace*{1cm} at\ }r=10h\mone\Mpc
\eea
In each case, $\calp_0$ is the \it linearly evolved \rm spectrum,
evaluated at the present epoch.

One sees that $\sigma_g(r)/b$ is the dispersion of the linearly evolved density
contrast with top hat filtering on the scale $r$, whereas $(2r\mthree
J_3(r))\half/b$ is its dispersion on the same scale with a different choice of
filter. As noted earlier, the dispersion of a filtered quantity evolves
linearly only as long as it is $\lsim1$, and this condition is only marginally
satisfied by the quantities encountered in the two conventional normalisation
schemes. The non-linear correction will be worse if $b\simeq1$ than if
$b\simeq 2$, and Couchman and Carlberg (1992) have claimed on the basis of
numerical simulations that it is indeed significant in the former case.
Evaluating the integrals, one finds that the $J_3$ normalisation is typically
10\% lower than the normalisation using $\sigma_g$, across the range of $n$ in
which we shall be interested. These are shown in figure 2. As remarked above,
different choices of transfer function parametrisation can give a correction
of around 10\% as well.

In what follows we always adopt the $\sigma_g$ normalisation scheme, in the
sense that normalisation of the spectrum is specified by quoting the quantity
$b$ \it defined \rm by \eq{sig}. In words, $b\mone$ is defined as the present
value of the dispersion of the \it linearly evolved \rm matter density
contrast, with top hat filtering on the scale $R_f=8h\mone\Mpc$. Only
occasionally do we invoke the much stronger assumptions, that the galaxy
correlation function is given by $\xi_{gg}(r)\simeq b^2\xi(r)$ and that
$\xi(r)$ is equal to the linearly evolved quantity on scales $r\simeq
10h\mone\Mpc$. Our most significant results, summarised in figure 11 later in
the paper, do not involve these assumptions.

Figure 3 shows $b\calp\half_0(k)$ for a selection of $n$ values, where as
always $\calp$ is the linearly evolved quantity and the subscript zero
indicates evaluation at the present epoch. The spectra for different $n$
typically cross at a scale of around $k^{-1}= 2\pi/32 h^{-1}$ Mpc. Figure 4
shows the corresponding dispersion $b\sigma(M)$ for $n=1$ and $n=0.7$, for
both Gaussian and top hat filtering. The mass scale $M$ runs from
$M=10^6\msun$, the Jeans mass at decoupling (Peebles 1980, Kolb \& Turner
1990) to $M=10^{18}\msun $, the mass of the universe on the scale of the CfA
survey. The comoving wavenumber $k\mone$ runs over the corresponding range of
filtering radius $R_f$.

For $n=1$ the spectrum $\calp(k)$ is practically flat on small scales. For
$n<1$ the power on small scales is reduced, though even for $n=0$ not by as
much as in the standard hot dark matter model (White, Frenk \& Davis 1983) or
with string-seeded hot dark matter (Albrecht \& Stebbins 1992)). As a result
the spectrum starts to fall on scales $k\mone\lsim .1\Mpc$. The implication of
this change for small scale structure is discussed in Section \ref{NONLIN}.

\section{Large Scales: Large Angle Microwave Background Anisotropies}
\label{MWB}
\setcounter{equation}{0}
\def\theequation{\thesection.\arabic{equation}}

Measurements of the cosmic microwave background anisotropy at large angular
scales provide the only evidence we possess as to the amplitude of the
spectrum on scales $\gsim 100h\mone \Mpc$. Ignoring the possibility of
reionisation, the surface of last scattering for the cosmic microwave
background lies some $190 h^{-1}$ Mpc inside the particle horizon
(Hogan, Kaiser \& Rees 1982), whose present distance is $2H_0^{-1}=6000h^{-1}$
Mpc. This means that an angular scale $\theta_0$ corresponds to a linear scale
$r=101\theta h\mone\Mpc$.

For angular scales $\gg 1^0$, the linear scale is much bigger than the horizon
size at last scattering. This means that the microwave anisotropy does not
involve such awkward physical processes as Thomson scattering from moving
electrons and the like, which vastly complicate matters when one reaches
scales of a degree or less (Bond \& Efstathiou 1987). It is in fact given by
the well-known and understood Sachs-Wolfe effect (Sachs \& Wolfe 1967), which
corresponds simply to fluctuations in the gravitational potential across the
microwave sky caused by the density fluctuations.

If re-ionisation occurs the distance of the surface of last scattering will be
multiplied by some factor $\epsilon<1$, and the Sachs-Wolfe formula will be
valid only on scales $\gg \epsilon\mone$ degrees. Re-ionisation is not thought
to be likely in the standard CDM model, due to the relatively late onset of
nonlinear behaviour which is further exacerbated with power-law spectra.
{}From now on the possibility of re-ionisation is ignored.

The only data we consider here are those from the COBE experiment (Smoot {\it
et al} 1992), which to date provides the only positive detection of
anisotropies. It corresponds to an angular resolution of order $10^0$, where
one may safely ignore the possibility of re-ionisation. The corresponding
linear resolution is of order $10^3h^{-1}\Mpc$. Experiments with smaller
angular
resolution have so far reported only upper limits, which are compatible with
the COBE detection for the CDM model with $n\lsim 1$.

Let us first review the pertinent results from COBE (Smoot {\it et al} 1992).
The most useful and significant measurement made is that of the variance of
the temperature fluctuations smoothed on a scale of $10^0$. The value quoted
is $(1.1 \pm 0.2) \times 10^{-5}$. Like all the COBE results, the error limits
quoted are $1$-sigma. Of less theoretical importance is the measurement of the
quadrupole anisotropy $Q_2$, which is also susceptible to experimental
contamination from the galaxy in particular. This is quoted as $Q_2 = (5 \pm
1.5) \times 10^{-6}$.

COBE also have results on all intermediate scales, readily obtained by the
appropriate smoothing of their basic data set. These are the data used to
calculate the $1$-sigma limits on the best fit primeval spectral slope $n =
1.1 \pm 0.5$. Unfortunately, this intermediate data has not been given in the
paper, and so consequently we must work solely with the information given
above and the slope limits.

Now we proceed to the formalism, which follows closely that of Scaramella and
Vittorio (1990). [See also Abbott \& Wise 1984a; Bond \& Efstathiou 1987;
Scaramella \& Vittorio 1988; Efstathiou 1990, 1991.] For studies at large
angles, the most convenient approach is the expansion of the temperature
fluctuation field in spherical harmonics. Dropping the dipole term which is
dominated by our peculiar velocity relative to the comoving rest frame, one
writes
\begin{equation}
\frac{\Delta T}{T} ({\bf x}, \theta, \phi) = \sum_{l=2}^{\infty}
	\sum_{m=-l}^{+l} a_l^m({\bf x}) Y_m^l(\theta, \phi)
\end{equation}
where $\theta$ and $\phi$ are angles on the sky and ${\bf x}$ is the observer
position. With gaussian statistics for the original density field, the
coefficients $a_l^m({\bf x})$ are gaussian distributed random variables of
position, with zero mean and with a rotationally invariant variance which
depends only on $l$,
\begin{equation}
\langle a_l^m({\bf x}) \rangle = 0 \; \; \; ; \; \; \; \langle |a_l^m({\bf
	x})|^2 \rangle = \Sigma_l^2
\end{equation}
where the angle brackets are averages over all observer positions. It is
important to appreciate however that what one can actually measure is only the
temperature anisotropies at a single observer point, our own. Thus in practice
one only gets a single realisation of the probability distribution functions
given above. This leads to statistical uncertainties, as an individual
realisation may not reflect the properties of the ensemble averaged system.
This effect, normally called the {\em cosmic variance}, is significant for
observables which depend only on a limited number of the $a_l^m$, and in
particular for the quadrupole which only depends on the five $a_2^m$ (two of
which are redundant rotational information). This is the fundamental reason
why the quadrupole measurement is not particularly useful for constraining
theories. On the other hand, the variance at $10^0$ depends on a significant
number of the $a_l^m$ and is far less susceptible to statistical vagaries.

The multipoles $Q_l$ as seen from a particular observer point are defined from
the variances of the individual $a_l^m$ as
\begin{equation}
Q_l^2 = \frac{1}{4\pi} \, \sum_{m=-l}^{+l} |a_l^m|^2
\end{equation}
In fact, the temperature autocorrelation function measured by a single
observer
\begin{equation}
C(\alpha) \equiv \left\langle \frac{\Delta T}{T} (\theta_1, \phi_1) \;
	\frac{\Delta T}{T} (\theta_2, \phi_2) \right\rangle_{\alpha}
\end{equation}
where the angle brackets indicate an average over all directions separated by
an angle $\alpha$, is given exactly in terms of the $Q_l^2$ via
\begin{equation}
C(\alpha) = \sum_{l=2}^{\infty} Q_l^2 \, P_l (\cos \alpha)
\end{equation}
Note that our $Q_l^2$ have a different normalisation to that of Scaramella and
Vittorio (1990). It is defined here so that $\sqrt{Q_2^2}$ is
exactly the quantity COBE measures.

Since the $Q_l^2$ are the sums of squared gaussian random variables, they are
described by chi-squared distributions with the appropriate number of degrees
of freedom; for the $l$-th multipole there are $2l+1$ degrees of freedom, and
in particular the quadrupole is described by a $\chi_5^{\, 2}$ distribution.

The expectation of $Q_l^2$, again averaged over observer positions, is
directly related to the variance $\Sigma_l^2$ of the $a_l^m$, via
\begin{equation}
\langle Q_l^2 \rangle = \frac{1}{4\pi} \, (2l+1) \Sigma_l^2
\end{equation}
The values of the $\Sigma_l^2$ are readily calculated. On these large angular
scales, the temperature anisotropies from the density perturbations are
dominated by the Sachs-Wolfe effect (Sachs \& Wolfe 1967). The appropriate
formula for this curvature contribution is (Peebles 1982)
\begin{equation}
\Sigma_l^2 ({\rm scalar}) = \pi \int_0^{\infty} \frac{{\rm d}k}{k} \, j_l^2
	\left( 2k/aH \right) \; \delta^2_H
\end{equation}
where we write `scalar' to distinguish from other contributions to the
variances. Here $j_l$ is the spherical Bessel function. With
$\delta_H^2(k)\propto k^{n-1}$ this becomes
\be \Sigma_l^2(\text{scalar})=\frac18 \left[\frac{\sqrt\pi}{2}l(l+1)
\frac{\Gamma((3-n)/2)}{\Gamma((4-n)/2)}
\frac{\Gamma(l+(n-1)/2)}{\Gamma(l+(5-n)/2)}
\right]
\frac{2l+1}{l(l+1)}\delta^2_H(H_0/2) \ee

As discussed in section \ref{INFL}, there is in power-law and extended
inflation models an additional contribution to the multipoles from
gravitational waves. The contribution to the squared multipoles from
gravitational waves is adequately given (for the range of $l$ values we will
use) from the scalar contribution above just by
\begin{equation}
\Sigma_l^2 ({\rm grav}) = \frac{12}{p} \, \Sigma_l^2 ({\rm scalar})
\end{equation}
so that
\begin{equation}
\Sigma_l^2 ({\rm total}) = \frac{15-13n}{3-n} \, \Sigma_l^2 ({\rm scalar})
\end{equation}
In contrast, natural inflation has a negligible gravitational wave
contribution, even though it can give a tilted spectrum just like the
power-law case.

\subsection{The quadrupole anisotropy}

As we shall see below, the quadrupole is not particularly useful for
constraining the models, and we shall ultimately drop it in favour of the
$10^0$ result. Nevertheless, it is worth examining the predictions of the
model to see why we come to this conclusion.

The above equations readily allow us to calculate the mean of the
predicted distribution function for $Q^2_2$.
It is given in figure 5, again
as a function of spectral slope, for both types of inflationary model. Because
the spectral normalisation is uncertain by the bias factor, we have plotted
the quantity $b^2 \langle Q_2^2 \rangle$. The COBE result of $2.5 \times
10^{-11}$ corresponds almost exactly to the mean prediction for the flat
spectrum when the bias is one, whereas at $n = 0.6$ the predicted mean is
nearly 12 times the COBE result even for the natural inflation scenario,
seemingly requiring a bias of over three. However, that conclusion neglects
the statistical nature of the quadrupole prediction.

As discussed above, the quadrupole prediction is not for a unique value but for
a $\chi_5^2$ probability distribution function ({\em pdf}) with the given
mean. For a given observer, the prediction is for a given realisation from
that distribution. Let us discuss the predictions in terms of a quantity $q
\equiv 10^{10} \, b^2 \, Q_l^2$. Then the mean value $\bar{q} = 10^{10} \, b^2
\, \langle Q_2^2 \rangle$, and the $\chi_5^2(q)$ {\em pdf} with that mean is
\begin{equation}
\chi_5^2 (q) = \frac{1}{3 \sqrt{2\pi} \left(\bar{q}/5 \right)^{5/2}} \,
	q^{3/2} \exp \left( - \frac{5q}{2\bar{q}} \right)
\end{equation}
This distribution has a very broad spread (the variance being $2/5$ of the
mean squared), and thus realisations of it may differ considerably from the
mean. The spread is indicated in figure 5 by the vertical bars through the
mean predictions.

Let us for the time being assume the COBE measurement to be perfect ({\em ie}
not subject to observational errors). Then one can only exclude a given theory
on the basis of some exclusion level, where the experimental result is far
along the tail of the distribution. A conventional choice would be that 95\% of
the distribution should predict values above (or below) the experimental
measurement. Thus one can exclude at 95\% confidence only those theories which
predict a mean sufficiently high that 95\% of the distribution is above the
COBE value of $q_{\rm exp} = 0.25 \, b^2$. This one can readily calculate via
error functions, to discover that one can only say with 95\% confidence that
$\bar{q} < 1.1 b^2$, corresponding to $\sqrt{\langle Q_2^2 \rangle} < 1.05
\times 10^{-5} b$.

One can loosen this constraint even further by incorporating the fact that the
COBE results possess experimental errors. One can model the COBE data with a
probability distribution function describing the expected values were the
experiment to be repeated --- a sensible choice might be to assume $Q_2^{{\rm
exp}}$ to be gaussian distributed with mean $5 \times 10^{-6}$ and width $1.5
\times 10^{-6}$ --- and use this to construct a {\em pdf}~~$p_{\rm exp}
(q_{\exp})$ for the experimental measurement $q_{\rm exp}$. One then tests
each member of the experimental distribution for exclusion and takes a
weighted mean, thus constructing a rejection functional on the predicted
theoretical {\em pdf}s as
\begin{equation}
{\cal R} = \int_0^{\infty} p_{\rm exp} (q_{\rm exp}) \left| 1-2 \int_0^{q_{\rm
	exp}} \chi_5^2(q) dq \right| dq_{\rm exp}
\end{equation}
where $|$ signals the modulus. Defined on one side of the probability
distribution like this, a value ${\cal R} > 0.9$ signals a 95\% rejection of
the theoretical {\em pdf} in the light of the modelled experimental data.
With the experimental modelling as suggested above, then predicted mean values
of $\bar{q}$ up to $2.3b^2$ are allowed, corresponding to 95\% confidence that
$\sqrt{\langle Q_2^2 \rangle} < 1.5 \times 10^{-5} b$.

Hence one sees that the statistical uncertainties in the theoretical
quadrupole prediction make it of little use in constraining theories. For
example, with this last constraint $n = 0.6$ is allowed with a mild bias of
1.15.

\subsection{The variance at $10^0$}

Following the filter function formalism of Bond and Efstathiou (1987) (see
also Efstathiou 1991), the mean (over all observer points) of the anisotropy
seen by a given experiment is given by
\begin{equation}
\left\langle \left( \frac{\Delta T}{T} \right)^2\right\rangle = \frac{1}{2\pi}
	\sum_l (2l+1) \Sigma_l^2 F_l
\end{equation}
where $F_l$ is a filter function appropriate to a given experimental
configuration.

For ground-based experiments which typically feature two and three beam
configurations, this filter function can be rather complex. For the COBE
experiment it is much simpler, as COBE reconstruct the smoothed fluctuation
field across the entire sky and calculate the variance of that. The original
COBE beam is well approximated by a gaussian with Full Width Half Maximum of
$7^0$. However, before calculating the variance they smooth again by
convolving with a further $7^0$ FWHM gaussian, a procedure equivalent to an
original smoothing by a $10^0$ FWHM gaussian. Such a gaussian has a variance
$\sigma^2 = (4.25^0)^2$. Thus the appropriate form of the filter function is
\begin{equation}
F_l = \frac{1}{2} \exp \left( - \left(4.25 \pi l/180 \right)^2 \right)
\end{equation}

It is now trivial to calculate the predicted $10^0$ variance as a function of
$n$, and in figure 6 we plot its square root, henceforth denoted $\left.
\frac{\Delta T}{T}\right|_{10^0}$, multiplied by the bias, for both power-law
and natural inflation. In the latter case, one sees that the results are
essentially exactly linear. We have been unable to show why this should be
analytically. The appropriate fitting function is
\begin{equation}
\left. \frac{\Delta T}{T} \right|_{10^0}(n) = \exp \left(2.62 (1-n) \right)
	\; \left. \frac{\Delta T}{T} \right|_{10^0} (n=1)
\end{equation}
and so for power-law inflation one has
\begin{equation}
\left. \frac{\Delta T}{T} \right|_{10^0} (n) = \sqrt{\frac{15-13n}{3-n}} \exp
	\left(2.62 (1-n) \right) \; \left. \frac{\Delta T}{T} \right|_{10^0}
	(n=1)
\end{equation}

One can also readily calculate the cosmic variance, giving the spread in
values about these means that would be measured by differently positioned
observers. The variance of the $Q_l^2$ is $2\langle Q_l^2 \rangle^2/(2l+1)$.
For the $10^0$ result, this is 10\% at $n=1$, rising to 12\% at n=0.6 (this
result remains true whether or not there is a gravitational wave contribution
to the anisotropies). This is a negligible correction to the larger COBE error
in the present observations, but is ultimately a limiting obstacle to the
conclusions one can draw on this large angular scale.

\subsection{Microwave background constraints}

In establishing constraints on the bias at fixed $n$, it is the $10^0$ data
which are of primary interest. Some particular values worthy of note are that
for the flat spectrum of $\left. \frac{\Delta T}{T}\right|_{10^0} = 1.05
\times 10^{-5}/b$, and that for $n=0.6$ of $\left. \frac{\Delta T}{T}
\right|_{10^0} = 5.17 \times 10^{-5}/b$ for power-law inflation and $\left.
\frac{\Delta T}{T} \right|_{10^0} = 2.99 \times 10^{-5}/b$ for natural
inflation. These of course are to be compared with the COBE observations of
$\left. \frac{\Delta T}{T}\right|_{10^0} = (1.1 \pm 0.2) \times 10^{-5}$.

The mean quadrupole prediction is also very similar to the flat spectrum
prediction with bias one, but the statistical uncertainties make this
comparison rather less relevant. It is worth recalling that an extrapolation
of the quadrupole from the COBE data at smaller angles, assuming a power-law
spectrum, gives a somewhat larger prediction for the {\em mean} quadrupole of
$16 \pm 4 \mu K$, corresponding to $(6 \pm 1.5) \times 10^{-6}$ (Smoot {\it
et al} 1992). Note that this extrapolation estimates the mean quadrupole, not
a specific realisation. Hence this again favours values of the bias not much
exceeding $1$, unless one inserts large scale power to boost the theoretical
prediction for the mean.

Lacking full access to the COBE data, one must make an operational choice as
to what to take as the observational limits. The COBE team provide fits of
their data to power-law spectra, where both the slope and amplitude are
treated as fitting parameters. The two pertinent pieces of information are
that the observed (root of the) variance at $10^0$ is $(1.1 \pm 0.2) \times
10^{-5}$, and the allowed slopes in the fit carried out by the COBE team are
$n = 1.1 \pm 0.5$, where both errors are $1$-sigma. These are of course not
independent, as $10^0$ is at the lower end of the fitting range. Nevertheless,
if one takes the upper $2$-sigma value of the extrapolated quadrupole ($24
\mu K$), then this actually gives a prediction for the $10^0$ variance which
is well above the $2$-sigma limit on the $10^0$ data (for $n=1$, the root of
the mean $10^0$ variance is about 2.03 times the mean quadrupole, the relative
factor falling as $n$ decreases). This conclusion (also implicit in
Efstathiou, Bond and White (1992)), argues that the fits are not rigid enough
to be of much use in constraining theories. Because of the cosmic variance and
also experimental uncertainties, the quadrupole itself also appears of little
use in this type of analysis. Consequently, we choose to adopt simply the
$10^0$ result as it stands, without further incorporation of the COBE data. We
relax their error bars to $2$-sigma, and for many purposes we are only
interested in the upper limit thus given of $1.5 \times 10^{-5}$. This is a
particularly useful way of utilising the results, as it seems likely that this
number can only go downwards if it is to avoid conflict with other experiments
and so any limits quoted on its basis are likely only to become stronger with
improved observations.

The full limits on the bias as a function of $n$ from this criterion, for
both styles of inflationary model, are plotted in figure 11 at the end of
section \ref{CLUSTER} along with other constraints derived in that section.
Some sample results are that for $n=1$ we have the obvious $0.7 < b < 1.5$,
while for $n=0.6$ we require $3.4 < b < 7.4$ (power-law inflation) and $2.0 <
b < 4.3$ (natural inflation). Let us finally recall the uncertainties of
normalisation. Were one to have used the $J_3$ spectral normalisation rather
than $\sigma_8$, these bias values would be lowered by about 10\%; the use of
the Bardeen {\it et al} (1986) transfer function rather than the one we use
would reduce the required bias by a further 10\%, and a higher baryonic
content raise it by a similar amount.

\section{Intermediate Scales: Galaxy Clustering and Bulk Velocity Flows}
\label{CLUSTER}
\setcounter{equation}{0}
\def\theequation{\thesection.\arabic{equation}}

Now we consider scales which are big enough that the filtered density contrast
is still evolving linearly, yet small enough that there is information about
the galaxy correlation function and about the peculiar velocity field. Thus we
are discussing scales from around $10h^{-1}$ Mpc up to perhaps $100h^{-1}$
Mpc. The galaxy correlation function is sensitive to the combination $b
\delta(M,\bfa x)$, whereas the peculiar velocity field is is sensitive to
$\delta(M,\bfa x)$ itself.

\subsection{The galaxy correlation function}

There is ever increasing evidence that the amount of large-scale galaxy
clustering is greater than can be accommodated in the standard CDM
cosmogony.
The most striking piece of evidence is provided by the measurement of the
galaxy angular correlation function $w(\theta)$ in the APM survey (Maddox {\it
et al} 1990, 1991), based on a sample of over two million galaxies. Further
evidence pointing to the same conclusion has been provided by the `counts in
cells' of the QDOT survey (Saunders {\it et al} 1991), and more recently by a
survey of redshifts of APM galaxies (Dalton {\it et al} 1992), and in the power
spectrum inferred from the CfA survey (Vogeley {\it et al} 1992) and that
inferred from the Southern Sky Redshift Survey (Park, Gott \& da Costa 1992).

Our earlier work with Will Sutherland (LLS) on structure from the power-law
inflation model was primarily concerned with an investigation of whether
or not this excess clustering could be intrinsic to the primeval spectrum,
were the spectrum to be of power-law form. In this subsection we recap briefly
on that analysis and its conclusions. The remaining results of that paper,
concerning microwave anisotropies and the formation of nonlinear structure,
are superceded by the other sections of this paper.

On the large angular scales where the discrepancy arises, the angular
correlation function $w(\theta)$ can readily be calculated in linear theory.
It is given (Peebles 1980) by Limber's equation as an integral over the galaxy
correlation function $\xi_{gg}(r)
= b^2 \xi(r)$, where the mass correlation
function $\xi(r)$ is defined in section \ref{SPECTRUM}. A linear calculation
is expected to be reliable on scales above about $2^0$. The results appear in
figure 7, reproduced from LLS. The flat spectrum $n=1$ falls well below the
observational data, but one can see that as the spectrum is tilted, the extra
power does indeed make itself evident in the clustering statistics. It is
suspected that the APM survey may contain small residual systematics which
bias the observational estimates upwards, so it seems reasonable to regard
values of $n$ from $0.3$ to $0.6$ as good fits to the excess clustering data.
(This is actually rather more restrictive than the range allowed by
Efstathiou, Bond and White (1992), as we see below.) We emphasise once more
that this result is independent of the value of the bias parameter.

It has become a common reference point to consider spectra with low effective
$\Omega h$ when considering clustering data. Figure 8 compares the processed
spectra for $n = 1$, $0.6$ and $0.5$ with that of a CDM model with a flat
spectrum but with $\Omega h \sim 0.2$. Such a model has been much touted,
especially by Efstathiou and collaborators (Efstathiou, Sutherland \& Maddox
1991; Efstathiou 1991; Efstathiou, Bond \& White 1992), as providing a good
fit to all of the excess clustering data. One sees that the tilted power-law
spectra resemble this spectrum much more closely than does standard CDM,
though with some additional reduction in short scale power. The $\Omega h \sim
0.2$ model has also been claimed as a good fit to the distribution of rich
clusters (Lilje 1992; Scaramella 1992), to the CfA survey (Vogeley {\it et al}
1992) and to the Southern Sky redshift survey (Park, Gott \& da Costa 1992),
all of which appear inconsistent with standard CDM. We thus expect tilted
spectra to also do well on these criteria, though we have not attempted a
direct comparison here as we lack suitable data.

One can quantify this connection somewhat by utilising a quantity introduced
by Wright {\it et al} (1992) called the excess power, which is a functional of
the power spectrum defined as
\begin{equation}
E[\calp] = 3.4 \, \frac{\sigma(25h^{-1} {\rm Mpc})}{\sigma(8h^{-1} {\rm Mpc})}
\end{equation}
For standard CDM and our transfer function, $E = 0.95$ (there is a very weak
dependence on the choice of transfer function, which still plays a role on
these scales). We find a reasonable fit is given simply by
\begin{equation}
E[n] = 1.44 - n/2 \; \; \; \; ; \; \; \; \; 0.3 \leq n \leq 1
\end{equation}
This can be related to the $\Gamma$ parameter of Efstathiou, Bond and White
(1992) as
\begin{equation}
\Gamma = \frac{1}{2} \left( \frac{1.88}{2.88-n}\right)^{10/3}
\end{equation}
The range of $\Gamma$ these authors consider a reasonable fit to the APM data
is $0.15 < \Gamma < 0.30$, corresponding to $0.15 < n < 0.67$, which we see is
actually rather looser than the range we took above.

The relation between $n$ and $\Gamma$ provides an excellent means to compare
results of different authors on intermediate length scales. It is worth
remembering though that well away from the scales $8h^{-1}$ and $25h^{-1}$ Mpc
used to define excess power the spectra do show significant differences, and
so this relation is not as useful for considering the microwave background
anisotropies, and even less so on short scales where the power-law spectra show
considerably less power than their $\Gamma$-equivalents.

It would be invidious of us to close this subsection without reference to a
huge body of literature which explains the excess clustering by a variety of
means other than by adding intrinsic power to the primeval spectrum. One way
in which this can be done is to modify the transfer function. Introducing a
cosmological constant with $\Omega_{\Lambda} \sim 0.8$ provides a realisation
of the $\Gamma \sim 0.2$ standard CDM model discussed above (Efstathiou,
Sutherland \& Maddox 1990; Efstathiou, Bond \& White 1992), which remains
viable though under threat from observations of quasar lensing (see {\it eg}
Fukugita {\it et al} 1992). A $17$ keV neutrino with a lifetime of the order
of 10 years can also provide an appropriate modification of the transfer
function (Bond \& Efstathiou 1991), though the evidence for such an object
from particle physics seems to be evaporating. Hybrid models combining both
hot and cold dark matter in carefully chosen combination are also finding
increasing favour (Schaefer \& Shafi 1992; Davis, Summers \& Schlegel 1992;
Taylor \& Rowan-Robinson 1992).

An alternative strategy is to modify the galaxy formation process itself. A
particularly elegant proposal is to allow quasars to suppress local galaxy
formation (Babul \& White 1991), though this must now contend also with the
COBE measurements implying a bias of around $1$. Cooperative galaxy formation,
a phenomenological (at present) model in which galaxy formation is favoured in
the neighbourhood of other galaxies, has also been proposed (Bower {\it et al}
1992), which can have the effect of a scale-dependent bias. Finally, it has
been suggested that a detailed analysis of nonlinear evolution assuming a bias
of around $1$ (made more attractive by COBE) coupled with a velocity antibias
shows that there is no conflict between the predictions of this model and the
observed galaxy clustering (Couchman \& Carlberg 1992).

\subsection{Bulk peculiar velocities}

There are many controversial aspects to the measurements of bulk velocity
flows, but nonetheless they provide a useful measure of the absolute magnitude
of the power spectrum on intermediate scales (for a recent review, see
Kashlinsky \& Jones (1991)). Quite detailed work on the bulk velocity flows
has been carried out for the power-law spectrum (and also allowing $\Omega <
1$) by Tormen, Lucchin and Matarrese (1992) and by Tormen {\it et al} (1992),
though our analysis will take a somewhat different and simplified approach.

Just as one can filter the density contrast, one can also filter the peculiar
velocity field to obtain a bulk flow quantity, on any desired scale. As long
the density contrast on that scale is in the linear regime, the bulk flow can
be constructed from the linearly evolved Fourier components $\bfa v_{\bf k}$,
given by (Peebles 1980; Kolb \& Turner 1990)
\be \label{VK}
\bfa v_{\bf k}=i \hat{\bf k} \frac{aH}{k} \delta_{\bf k} \ee
where $\hat{\bf k}$ is the unit vector in the \bfa k-direction.

Through the galaxy redshifts, the peculiar velocity field at the present epoch
is observable within a sphere around us whose radius is a few hundred Mpc.
Smearing over a scale $\gsim 10h\mone\Mpc$, one obtains a bulk velocity field
which is still evolving linearly. Taking the mass density contrast as being
$b\mone$ times the galaxy number density it too can be observed, and inserting
both quantities into \eq{VK} the value of $b$ can be estimated. This method of
estimating $b$ has been extensively explored for the standard case $n=1$.
Until recently, applied for instance to the region around the Virgo
supercluster, it seemed definitely to require a bias factor $b$ significantly
bigger than 1 (Peebles 1980; Kolb \& Turner 1990). With the advent of better
measurements this is no longer the case, and a value $b=1$ is apparently
possible (Dekel 1991). Without direct access to the data it seems difficult to
know how this result is affected if $n<1$. The possibility of a velocity bias
(Carlberg, Couchman \& Thomas 1990; Couchman \& Carlberg 1992) further
complicates the analysis.

We base our comments here around the recent results concerning IRAS galaxies
form the QDOT survey. Our analysis essentially mimics that of Efstathiou,
Bond and White (1992). One gets two useful statistics from the QDOT survey.
The first is that, in combination with bulk flow data, one can estimate
directly the bias of IRAS galaxies. In general, the bias of the infra-red
selected (and thus typically young) IRAS galaxies will not be the same as that
of the optically selected galaxies discussed thus far, and so we denote this
bias by $b_I$. It is often stated that IRAS galaxies are somewhat less
clustered than their optical counterparts; for example Saunders,
Rowan-Robinson and Lawrence (1992) suggest $b_I = (0.69 \pm 0.09)b$ at the
$1$-sigma level. According to Taylor and Rowan-Robinson (1992), there are
three independent dynamical estimates of $b_I$, giving respectively $b_I =
1.23 \pm 0.23$, $1.16 \pm 0.21$ and $1.2 \pm 0.1$ where all errors are
$1$-sigma. One cannot statistically combine errors from the same data set, so
we adopt as a result (which we assume has something like $2$-sigma errors)
that $b_I = 1.2 \pm 0.3$.

The second useful statistic is the dispersion of counts in cells in $30 h^{-1}$
Mpc cubes, given as $0.46 \pm 0.07$ (Saunders, Rowan-Robinson \& Lawrence
1992) where we have doubled the quoted error bars. Because we have a good
estimate of the bias, we can immediately calculate the mass variance in these
cubes, and then use the spectrum to calculate the variance at any other scale.
In particular, one can scale down to $8 h^{-1}$ Mpc and thus determine limits
on the bias parameter $b$. Note that the use of the optical bias parameter is
just a way of representing the results, without implying any reference
either to optically selected galaxies or to scales on the point on
nonlinearity. We note also that even if one did not know the variance of the
QDOT counts in cells, one would already have strong limits from the knowledge
that optical galaxies are not too much more clustered than IRAS ones.

The comparison between the bias limits obtained by this method and those
obtained from the COBE measurements represent a particularly strong
combination. In our earlier paper (Liddle \& Lyth 1992) we pointed out that
because of the strong gravitational contribution to the microwave
anisotropies, this comparison rules out at least the simpler models of
extended inflation because of the $n < 0.75$ bubbles constraint on these
models. In figure 11 at the end of this section, these constraints appear
together and highlight the allowed regions of the $n$--$b$ parameter plane.

Before finishing this section, we mention one other simple comparison which
can be made --- to the bulk flows in spheres as measured by POTENT. This
approach considers the dispersion of $v=|\bfa v|$, given on each scale through
the spectrum of $v$,
\begin{equation}
\calp_v(k) = \rfrac{aH}{k}^2  \calp(k)
\end{equation}
Using this approach it is easy to consider the effect of $n$. Figure 9
illustrates the scaling of the dispersion with $R_f$ for a top hat filter.
Smaller values of $n$ give significantly higher predictions, but the scaling
with $R_f$ is roughly the same.

To compare this result with observation we use the velocity field
reconstruction results from the POTENT method, pioneered by Bertschinger,
Dekel and collaborators (Bertschinger \& Dekel 1989; Dekel, Bertschinger \&
Faber 1990; Bertschinger {\it et al} 1990; Dekel 1991). It is well known (Kolb
\& Turner 1990) that typical theories, including standard hot and biassed cold
dark matter models and also models seeded by topological defects, tend to
predict bulk velocities rather lower than those observed, particularly for
standard CDM with high bias. Nevertheless, it is not trivial to compare theory
with observations, because one predicts only the {\em rms} velocity averaged
over the entire universe and the observed distribution may not be a fair
sample. In particular, the possible contaminating effects of the great
attractor (if backside infall can be unambiguously identified (Mathewson, Ford
\& Buchhorn 1992)), may distort observations. In the Monte Carlo work analysing
bulk flows of Tormen {\it et al} (1992), special criteria for choosing the
observer points are employed {\em before} statistical comparisons are made.

The POTENT results are obtained via a two-stage smoothing (Dekel 1991). First,
the original data is smoothed with a gaussian of radius $12h^{-1}$ Mpc, and
then a top hat of radius $R_f$ is used to provide the published data. In
figure 10, we apply this two-stage smoothing to our spectra, which reduces the
predictions, especially at short scales. These are compared with the POTENT
data at different biasses. It is seen that standard CDM at bias $1$ performs
rather well, significantly better than $n=0.6$ at bias $2$. Nonetheless, it
appears that either model is still tenable given the statistical nature of the
quantity being observed. In particular, the influence of large scales filters
down significantly to small scales with this statistic, so an unexpectedly
large fluctuation (and the great attractor would certainly be one in these CDM
models) can easily move the whole observed data set to well above the mean, as
predicted from the spectrum, that would be observed in a fair sample.

Some numerical analysis (based on Monte Carlo simulations) of bulk velocity
flows with power-law spectra has been made by Tormen {\it et al} (1992), who
consider 18 models combining all possible combinations of $b = 1$, $1.5$, $2$;
$n=1$, $0.5$, $0$; $\Omega = 0.4$, $1$. Of most interest to us is the $\Omega
=1$, $n = 0.5$, $b = 1.5$ model, which most closely represents the criteria we
are establishing in this paper. They employ a maximum likelyhood test, and
find that this model is twice as likely as the model with flat spectrum and
bias $1.5$, and is four times more likely than the flat spectrum with bias
$1$.

\subsection{Summary of the intermediate/large scale results}

One of the key aims of this paper is to investigate the constraints in the
$n$--$b$ parameter plane. The most significant constraints from large and
intermediate scales are plotted in figure 11, and all the lines on this figure
are to be taken as representing the $2$-sigma range on either side of the
(unplotted) mean value. The constraints plotted are
\begin{itemize}
\item The upper limit on $n$ (independent of $b$) obtained from requiring a
fit to the APM data (dotted line). We have taken our constraint to be $n <
0.6$ (see figure 7), though we note that Efstathiou, Bond and White (1992)
allow a range for their $\Gamma$ which is roughly equivalent to letting $n$ be
as large as 0.67, which is very conservative.
\item The limits on bias as a function of $n$ from the QDOT survey
(dot--dashed lines), as discussed in section \ref{CLUSTER}.2. In a sense these
are the weak link in the diagram, as although we have been very conservative
in the errors on this observation it is not inconceivable that changes will be
seen in the future. Nonetheless, we believe that these limits should be taken
very seriously and may well strengthen.
\item The COBE limits on bias as a function of $n$ for power-law and extended
inflation (solid lines) and for natural inflation (dashed line), as discussed
in section \ref{MWB}.3. These indicate the upper and lower $2$-sigma range of
the COBE $10^0$ data. These are very strong results, as it seems inevitable
that if any change to the observations occurs it will be to lower the
fluctuations and thus increase the required bias.
\end{itemize}
By themselves, the QDOT and COBE data allow a region of the $n$-$b$
plane which is roughly a triangle with corners
$\{n,b\}$ = $\{.7,1.6\}$, $\{1,1.6\}$ and $\{1,1\}$.
 Leaving aside for the moment the problematical galaxy clustering
represented by the APM data, we ask in the next section whether
data on smaller scales can discriminate between points within this
triangle.

\section{Small Scales: Galaxies and Galaxy Clusters}
\label{NONLIN}
\setcounter{equation}{0}
\def\theequation{\thesection.\arabic{equation}}

Now we study small scales  $k\mone\lsim10h^{-1}\Mpc$, corresponding to $M\lsim
10^{15}\msun$. This is the range on which gravitationally bound objects exist,
in particular galaxies with mass $10^6\msun\lsim M\lsim 10^{13}\msun$, and
galaxy clusters (including the small ones known as groups) with mass
$10^{13}\msun\lsim M \lsim 10^{15}\msun$.

In contrast with the treatment so far, we shall not attempt in this section to
work at the $2$-sigma level. The uncertainties in observational data, where
they are considered at all, will be taken simply as those quoted in the
literature.

For small scales, there are many quantities which one might hope to calculate
and compare with observation. A partial list is the following.
\begin{itemize}
\item The redshift of formation of a given class of objects, such as bright
galaxies or rich galaxy clusters.
\item The number density $n(z,>M)$ of all gravitationally bound systems with
mass bigger than $M$, which exist at redshift $z$.
\item The mean virial velocity of a given class of objects.
\item The number density $n(z,>v)$ of
all gravitationally bound systems with virial
velocity bigger than $v$, which exist at redshift $z$.
\item The correlation function $\xi\sub{obj}(r)$ of a given class of objects.
If it has the same shape as the correlation function $\xi(r)$ of the mass over
scales of interest, it is useful to define a bias parameter by
$\xi\sub{obj}=b\sub{obj}^2 \xi$.
\item The angular correlation $w(\theta)$ of bright galaxies, on the angular
scales $\theta\lsim2^0$ which correspond to linear scales $r\lsim
10h\mone\Mpc$.
\item The dispersion $\sigma_v(r)$ of the relative velocity of a pair of
galaxies separated by distance $r$.
\end{itemize}

As galaxies and clusters are gravitationally bound objects, these quantities
cannot be calculated using only the linear cosmological perturbation theory
which was adequate for intermediate and large scales. Considerable insight
can, however, still be gained from the linear theory if it is combined with
order of magnitude estimates inspired by a spherical model of gravitational
collapse. The general conclusion from this approach is that the $n=1$ case is
viable, with $b$ somewhere in the range $1\lsim b\lsim 2$. In the context of
the present paper, we wish to ask whether the linear approach will also permit
other values of $n$ and $b$, within the triangle allowed by the QDOT and
COBE data.

In the $n=1$ case, small scales have also been studied through non-linear
calculations, namely $N$-body simulations. Most such calculations agree that
the $n=1$ case is viable on small scales (Davis \it et al \rm 1985; Gott \it
et al \rm 1986; White \it et al \rm 1987; Carlberg \& Couchman 1989; Frenk \it
et al \rm 1990; Carlberg \it et al \rm 1990; Bertschinger \& Gelb 1991;
Couchman \& Carlberg 1992), though
some find that it has problems (Suto \it et
al \rm 1992, and other references sited there).
 A useful summary of the situation for $n=1$ has been given by
Davis \it et al \rm (1992). The corresponding calculations for $n<1$ are so
far in their infancy (Vogeley {\it et al} 1992; Park {\it et al} 1992; Cen
{\it et al} 1992), but so far they seem to
suggest that a value $n<1$ is preferred. We will
not consider numerical simulations further in the present paper.

\subsection{Gravitational collapse}

The linear approach makes essential use of the filtered density contrast
$\delta(M,\bfa x)$. Following the usual practice for theoretical calculations,
we use the Gaussian filter, defined by \eqst{gone}{gtwo}.
the dispersion of the filtered density contrast is thus
\be \sigma(M)=\int^\infty_0 \calp(k)\exp(-R_f^2 k^2)\frac{dk}{k}
\label{gfdis} \ee

At this point we need to distinguish between the density contrast of
the cold dark matter, and that of the baryonic matter.
They are the same on filtering scales $M$ well in excess of $10^6\msun$.
As $M$ is reduced below this figure, the
density contrast of the baryonic
matter falls off sharply, because
pressure forces prevent its growth until long after decoupling
(Peebles 1980; Kolb \& Turner 1990).
In contrast the cold dark matter has negligible pressure,
and its density contrast
grows like $(1+z)\mone$ just as it does on larger scales.
This difference is presumed to account for the absence of luminous
galaxies with mass below $10^6\msun$. In what follows,
$\delta(M,\bfa x)$ will always denote the density contrast of the
cold dark matter, though we are usually interested only in
scales bigger than $10^6\msun$.

As long as it is less than $1$, the
filtered density contrast evolves linearly
except in those rare regions of space
where it exceeds $1$ in magnitude. In the 50\% of these regions where it is
positive, gravitational collapse takes place. At least initially, the
collapsing regions have mass $\gsim M$, because the filtered density
contrast does not `see' structure on smaller scales.%
\footnote
{As discussed later, this statement
needs modifying when the collapsing regions are very rare.}
In the approach that we are considering, gravitational collapse is modelled by
taking the collapsing region be spherically symmetric. One then finds (Peebles
1980, Eqs.~19.50 and 19.53) that when a given mass shell stops expanding, the
mean density inside it is a factor $9\pi^2/16$ times bigger than the mean
density of the universe. If the matter inside the shell had evolved linearly
(density contrast $\propto (1+z)\mone$), its density contrast at that time
would have been $\delta=(3/5)(3\pi/4)^{2/3} =1.06$. After it has stopped
expanding, the  shell collapses. If spherical symmetry continues to hold, and
one neglects pressure forces, the shell has collapsed to a point by the time
that the age of the universe has increased by precisely a factor 2,
corresponding to a mean density contrast within the linearly evolved shell of
$\delta=(3/5)(3\pi/2)^{2/3}=1.69$. Numerical studies (eg. Peebles 1970)
indicate that by about this time, pressure forces will in fact have virialised
the random motion of the constituents of the object. After this initial
virialisation, the object can lose energy (dissipation) which further
increases its virial velocity.

The use of the spherical collapse model in the present context
is somewhat problematical. First, filtering on
a mass scale $M$ will distort the
profile of a peak with mass of order $M$, making it lower and broader.
Thus the linearly evolved \it filtered \rm density contrast will not have
precisely the behaviour described above, even for a spherical peak. Second,
the departure from sphericity of a linearly evolved collapsing region
`seen' by the filtered density contrast
can be
estimated from the linear theory. As we shall see
it is quite significant, and that of the true (unfiltered) density
contrast will be more so.
Finally, any initial departure from sphericity is amplified during the collapse
(Peebles 1980).
Nevertheless, the fact that the peaks of the density contrast are at
least roughly spherical (during the linear regime), as well as
numerical simulations
that have been done for the $n=1$ case,
indicate that two features of the spherical collapse model
translate into
reality.
\begin{itemize}
\item The regions with mass $\gsim M$ which have
undergone gravitational collapse can be at least approximately identified with
the regions where the linearly evolved density contrast $\delta(M,\bfa x)$
exceeds some threshold $\delta_c$.
\item A collapsing region does not fragment into a large number of separate
objects, which means that the mass of the resulting gravitationally bound
systems is also $\gsim M$.
\end{itemize}

The optimal choice of the threshold $\delta_c$ is a matter of debate. Many
authors take the value $1.69$ inspired by the above spherical collapse model.
On the other hand, comparison of the linear estimate of $n(>M)$ (described
below) with the estimate from numerical simulations suggests a smaller value,
Carlberg and Couchman (1989) advocating $\delta_c=1.44$ and Efstathiou and
Rees (1988) advocating $\delta_c=1.33$
(but see also Brainerd and Villumsen (1992)).
 When making numerical estimates we set
$\delta_c=1.33$, while keeping in mind the effect of taking a bigger value.

As the collapsed regions $\delta(M,\bfa x)>\delta_c$ represent exceptionally
large fluctuations of a Gaussian random field, there are powerful mathematical
results concerning their stochastic properties. We shall use some of them in
what follows, drawing extensively on the work of Bardeen \it et al \rm (1986,
henceforth BBKS).

\subsection{The epoch of structure formation}

On each mass scale, the linear regime ends at
the epoch $z\sub{nl}(M)$ when
the filtered linearly evolved dispersion $\sigma(z,M)$ of the density contrast
is equal to 1. Since this quantity is proportional to $(1+z)\mone$, the epoch
is given by
\be 1+z\sub{nl}(M)=\sigma_0(M) \ee
where the subscript $0$ denotes the present value.
At this epoch, the formerly
rare regions where gravitational collapse is taking place on scales $\gsim M$
become common. Before studying the linear regime which is our main
concern, let us ask how this gravitational collapse process proceeds.
In particular we want to know whether the bottom-up picture of structure
formation, known to occur if $n=1$, will also occur if $n<1$.

At the epoch $z\sub{nl}(M)$ the collapsing regions, which have mass $\gsim M$,
give rise to gravitationally bound structures with mass $\gsim M$
provided that they do not fragment into a large number of pieces.
If $\sigma(M)$ is increasing significantly as the mass
decreases, then at this same epoch the density
contrast filtered on mass scales substantially bigger than $M$ is still
evolving linearly in most parts of the universe. In that case, only a small
fraction of the mass  of the universe is bound into objects with mass much
bigger than $M$. Moreover, $N$--body simulations
indicate that fragmentation does not then occur, so that the objects formed at
the epoch $z\sub{nl}(M)$ do not have mass much less than $M$ either. The
result is a bottom-up scenario, in which gravitationally bound systems of
successively bigger mass $M$
form at the successively later epochs $z\sub{nl}(M)$.
After a system has formed, it may become part of a bigger gravitationally
bound system, remaining a discrete object,
but it may also merge with other systems.

The scale dependence of $\sigma(M)$ depends on that of
the spectrum $\calp(k)$, on the corresponding scale
$k\mone\sim R_f$. In the cosmologically interesting
range $M>10^6\msun$, these
scale dependences were already shown
in figures 3 and 4 both for the standard choice $n=1$
and for
some smaller values. One sees that for $n=1$,
$\sigma(M)$ increases steadily as $M$ decreases, corresponding to the
fact that $\calp(k)$ does not turn over as $k\mone $ decreases.
In contrast, for $n<1$ the spectrum $\calp(k)$ turns over below some mass
scale,
so that $\sigma(M)$ levels out.
This difference in the small scale behaviour arises from the
small scale behaviour $\calp(k)\propto
T^2(k) k^{(3+n)}$. According
to the parametrisation \eq{tran}, $T^2(k)$ is proportional to $k\mfour$
which makes $\calp(k)\propto k^{n-1}$. This means that
$\sigma(M)$ increases logarithmically as $M$ decreases if $n=1$,
whereas if $n<1$ it approaches the unfiltered value $\sigma$,
the difference going to zero
like $k^{n-1}$.
(Strictly speaking the above statements are modified by
logarithmic factors because $T^2$
goes like $k\mfour\log^2 k$ (BBKS), but they do not  change the qualitative
behaviour.)

The small scale behaviour described in the last paragraph
is expected to persist down to some
very small scale, below which the spectrum cuts off sharply
and $\sigma(M)$ becomes practically equal to the unfiltered quantity
$\sigma$. This scale is called the coherence scale and depends on
the nature of the cold dark matter.
If the cold
dark matter consists of subnuclear particles, the coherence scale
is determined by their interactions. For instance, if it
consists of the lightest
supersymmetric particle the coherence scale is the Hubble scale
at the epoch when the particle becomes non-relativistic.
If it consists of the
axion, the coherence scale is the Hubble scale at the epoch
when the axion acquires mass through QCD effects.

On the basis of these considerations, we
arrive at the following picture of structure formation,
which covers the case $n<1$ as well as the familiar case
$n=1$. The first structure forms at some
epoch given by
\be 1+z\sub{nl}=\sigma_0 \ee
where $\sigma_0$ is the unfiltered quantity.
For $n=1$, this epoch is very early
and depends on the nature of
the cold dark matter. Decreasing $n$ makes the epoch  later,
and to an increasingly good approximation
allows it to be calculated without knowing the nature of the
cold dark matter.
After this epoch, structure forms according
to the bottom-up picture.

Let us estimate the maximum mass $M\sub{max}$,
above which structures collapsing at a given epoch $z\sub{nl}(M)$
are rare.
It should be such that $\sigma_0(M\sub{max})$ is
significantly less than 1, so that the density contrast filtered
on the scale $M\sub{max}$ is still evolving linearly
almost everywhere
in the universe.
To be definite, let us require that it
is still evolving linearly in 90\%
of the volume of the universe. From the Gaussian distribution, this
corresponds to $\sigma_0(M\sub{max})= 1/1.63=.61$.
For $M=10^6\msun$ this criterion gives
$M\sub{max}=10^{9}\msun$ ($10^{10}\msun$) for $n=1$ ($.7$).
For this value of $M$, the epoch
$z\sub{nl}(M)$ is a slowly varying function of $M$, so
the conclusion is that
a broad band of masses downwards of $M=10^9\msun$ to $10^{10}\msun$
collapse at the same time, around the epoch $z\sub{nl}
(10^6\msun)$.
For $M=10^{12}\msun$, the criterion gives $M\sub{max}=
2\times 10^{13}\msun$, more or less independent of $n$.
For this value of $M$, the epoch $z\sub{nl}(M)$ is a fairly rapidly
varying function, so the conclusion is that a fairly narrow band of
masses $M=10^{12}\msun$ to about $10^{13}\msun$ collapse
around the epoch $z\sub{nl}(10^{12}\msun)$.

We conclude that the bottom-up picture is more or less the same for
any $n$ in the range $.7<n<1$. A broad range of masses below about
$10^{9}\msun$ collapses at about the same time, but subsequent collapse
takes place in an increasingly narrow mass range.

The epoch $z\sub{nl}(M)$ can be read off figure 4, for
$n=1$ and $n=.7$. It becomes more recent as $M$ or $n$ are
reduced.
Let us consider three representative values
of $M$. First take
$M=10^6\msun$. Then $z\sub{nl}(M)$ is quite early,
$1+z\sub{nl}(M)=18/b$ ($9/b$) for $n=1$ ($.7$).
A broad range of mass scales downwards of about $10^9
\msun$ ($10^{10}\msun$) collapse at around this epoch.
Second, take $M=10^{12}\msun$. Then
$1+z\sub{nl}(M)=4.5/b$ ($3.7/b$) for $n=1$ ($.7$).
This is the epoch when a significant fraction of the mass of the
universe collapses into objects with mass of order $10^{12}\msun$,
the mass of large galaxies. However, the favoured explanation of a
bias factor $b>1$ is that \it luminous
\rm galaxies originate from exceptionally high peaks of the
evolved density contrast, and therefore form well before the
epoch $z\sub{nl}(M)$
(BBKS; Efstathiou 1990). Finally consider
$M=10^{15}\msun$. Then,
$1+z\sub{nl}(M)=.82/b$, almost independent of $n$.
This means that the epoch $z\sub{nl}(M)$, at which a significant
fraction of the mass collapses into large galaxy clusters,
lies in the future. The prediction is therefore that
the observed large clusters originated from exceptionally high peaks of the
density contrast, which again implies a bias factor for these objects
(Kaiser 1984). The bias factor for galaxies and galaxy clusters will
be discussed quantitatively below.

Finally, we consider a different criterion which
is sometimes proposed for the
epoch of structure formation, namely
that it is the epoch when the linearly evolved
spectrum $\calp(k)$ is equal to 1 on the corresponding scale $k\mone=R_f$.
This is practically equivalent to the criterion $\sigma(M)=1$ if $\calp(k)$ is
significantly increasing, say like $k^m$ with $m\sim 1$, because then the
filtered spectrum is $\calp(k)\exp(-k^2R_f^2)
\simeq\calp(R_f\mone)\delta(\log(kR_f))$, where $\delta$ denotes the
Dirac delta function.
 The condition for this to be the
case is more or less the same as the condition for bottom-up structure
formation to occur. It fails, in particular, when $k$ corresponds to the
maximum of $\calp$ which occurs when $n<1$, so it is not true that in that
case the first structure forms at the epoch when $1+z$ is equal to the maximum
value of $\calp(k)$, as was incorrectly stated in LLS.

\subsection{The statistics of the collapsed regions}

Now we study the stochastic properties of the collapsed regions, defined as
regions where the linearly evolved density contrast $\delta(M,\bfa x)$ exceeds
a threshold $\delta_c$. In them, $\delta(M,\bfa x)$ is more than $\nu$
standard deviations above zero, where
\bea \nu(z,M)\eqa \delta_c/\sigma(z,M) \label{nufirst} \\
\eqa \delta_c (1+z)/\sigma_0
(M) \\
\eqa \delta_c\frac{1+z}{1+z\sub{nl}(M)} \label{nudef} \eea
We are working in the linear regime, corresponding to $\nu
>\delta_c>1$.

The collapsed regions occupy a volume fraction $V$ given by the Gaussian
distribution,
\be \frac{{\rm d}V}{{\rm d}\nu}=\frac1{\sqrt{2\pi}} e^{-\nu^2/2}
\label{dgauss} \ee
leading to
\bea V(\nu)
\eqa\rm{erfc}(\nu/\sqrt2)/2\\
\eqa (2\pi)\mhalf\nu\mone e^{-\nu^2/2}
(1-\nu\mtwo+O(\nu\mfour) ) \label{gauss} \eea
For $\nu=1,2,3,4$ the volume fraction is $V=.16,.023,.0013,.000031$. In
practice one is not interested in values $\nu\gsim4$, because the collapsed
regions are then too rare to be physically significant. The corresponding mass
fraction is about $(1+\delta_c)=2$ to $3$ times bigger than the volume
fraction.

To say more one needs to know the shape of the spectrum. We shall list the
relevant results given by BBKS. They involve only two moments of the spectrum,
defined by
\bea
<k^{2}(M)>\eqa \sigma \mtwo(M) \int^\infty_0 k^{2} \exp(-k^2 R_f^2) \calp(k)
\frac{{\rm d}k}{k}
\\
<k^{4}(M)>\eqa \sigma \mtwo(M) \int^\infty_0 k^{4} \exp(-k^2 R_f^2) \calp(k)
\frac{{\rm d}k}{k} \eea
The quantity $\langle k^2 \rangle$ is the mean
of the $\nabla^2$ operator, ie., of
the quantity $\delta\mone\nabla^2\delta$.
Similarly, $\langle k^4 \rangle$ is the mean
of $\nabla^4$.

A relevant length scale is
defined by $R_*^2=3\langle k^2 \rangle/\langle k^4 \rangle$. For any
spectral index $n>-1$, it is easy to show that in the limit
of small filtering scale $R_f$,
\be \frac{R_*}{R_f}=\rfrac6{1+n} \half \ee
For the case of CDM with $.7<n<1$, the ratio is
in the range $1$ to $3$ for the entire range of cosmologically interesting
masses.

Another relevant length scale
is $\langle k^2 \rangle\mhalf$. On large filtering scales,
such that $\calp(k)$ is increasing fairly strongly at $k\mone\simeq R_f$,
the ratio $\langle k^2 \rangle\mhalf/R_f$ is close to 1.
As the scale is reduced it increases, but is $\lsim 10 $
in the cosmologically interesting range $M>10^6\msun$.

Finally, it is convenient to define the dimensionless parameter
\be \gamma(M)=\langle k^2 \rangle/\langle k^4 \rangle\half \label{gamm} \ee
It falls from about $.7$ to about $.3$ as $M$ decreases from $10^{15}\msun$ to
$10^6\msun$, for $.7<n<1$.

For sufficiently large $\nu$, each collapsed region is a sphere surrounding a
single peak of $\delta$. However, the departure from sphericity is
considerable in the cosmologically interesting regime. BBKS show that a
quantity $x\mone$, which is roughly the fractional departure from sphericity,
is well approximated by
\be
x=\gamma\nu+\theta(\gamma,\gamma \nu)
\ee
where
\be
\theta(\gamma,\gamma \nu)=\frac{3(1-\gamma^2)+(1.216-.9\gamma^4)
\exp[-\gamma/2(\gamma\nu/2)^2]}
{\left[3(1-\gamma^2)+.45+(\gamma\nu/2)^2\right]\half+\gamma\nu/2}
\ee
The asphericity is plotted in figure 12 for $M=10^8\msun$ and $10^{14}\msun$,
for both $n=1$ and $n=.7$, and is seen to be $\gsim .3$ even at $\nu=4$
and $10^{14} \msun$. We emphasise that this is the asphericity seen in
the linearly evolved, filtered density contrast. The asphericity in
the true, unfiltered density contrast will be bigger, and will increase
during collapse.

Three useful number densities are given by BBKS. First, the density $n_\chi$
of the Euler number of the surfaces bounding the collapsed regions is
\be
\frac12 n_\chi(\nu,<k^2>)= \frac{(\langle k^2 \rangle/3)\threehalf}{(2\pi)^2}
(\nu^2-1) e^{-\nu^2/2} \label{nchi}
\ee
Second, the number density of upcrossing points on these surfaces is
\be
n\sub{up}(\nu,<k^2>,\gamma)
=\frac{(\langle k^2 \rangle/3)\threehalf}{(2\pi)^2}
\left[ \nu^2-1+\frac{4\sqrt3}{5\gamma^2(1-5\gamma^2/9)\half}
\exp(-5\gamma^2\nu^2/18)\right] e^{-\nu^2/2} \label{nup}
\ee
An upcrossing point on a surface of constant $\delta$ is defined as one where
$\del \delta$ points along some chosen direction.

The third  number density is $n\sub{peak}$, the number density of peaks
which are more than $\nu$ standard deviations high. BBKS give
expressions for $n\sub{peak}$, but they point out also that in the
cosmologically interesting regime it is quite well approximated by
$n\sub{up}$. We shall use this approximation in what follows. It suggests that
if a collapsed region contains several peaks, they are not buried deep inside
it; rather, the boundary of the region is presumably corrugated, wrapping
itself partially around each peak.

In the limit $\nu\gamma\gg1$, $n\sub{peak}=n\sub{up}=\frac12 n_\chi$. This is
in accordance with the fact that in this regime, each surface is a sphere
surrounding a single peak. It contributes $+1$ to the number of upcrossing
points, and $+2$ to the Euler number. As $\nu$ decreases the surface becomes
deformed, but at first its contributions to $n\sub{up}$ and $n_\chi$ are not
affected. Eventually though, it may become so corrugated that it has more than
one upcrossing point, and may become a torus so that its Euler number is less
than $2$ (of course it then has more than one upcrossing point whatever its
shape). As a result, its contribution to $n_\chi/2$ becomes less than its
contribution to $n\sub{up}$. The ratio $2n\sub{peak}/n_\chi$
(approximated as
$2n\sub{up}/n_\chi$) is plotted in figure 13. (We are not interested in the
regime $\nu<1$, but for the record $n_\chi$ is negative there, indicating that
the surfaces $\delta(M,\bfa x)=\nu\sigma$ definitely do not have spherical
topology; in fact they percolate leading to a sponge--like topology for the
regions $\delta(M,\bfa x)>\delta_c$ (Melott 1990).)

One would like to know the number density $n\sub{coll}$ of the collapsed
regions. It satisfies the inequality $n_\chi/2<n\sub{coll}<n\sub{peak}$, and
so is equal to $n_\chi/2$ in the limit $\nu\gamma\gg1$. According to figure
13, this defines it within a factor 2 for $\nu>3$. The average number $N$ of
peaks per collapsed region satisfies $1< N < 2n\sub{peak}/n_\chi$, so
according to figure 13 it is no bigger than 2 for $\nu\gsim 3$.

\subsection{The mass of the collapsed regions}

As the filtered density contrast does not contain structure on scales much
less than the filtering scale $R_f$, we expect the average
radius of a peak to
be $\gsim R_f$. It can be estimated from the number density of peaks with
arbitrary height $n \sub{peak}(-\infty)$, which is equal to $.016R_*\mthree$
where the length $R_*$ was defined at the beginning of the last subsection.
This is also the number density of troughs, so a rough estimate of the
average
radius of a peak or trough at half height is
$(2n
\sub{peak}(-\infty)\mthird /4=.8R_*$.
As we discussed earlier, $R_*/R_f\simeq1$ to 3
for all cases of interest, so the average peak size is roughly of order $R_f$.
Since there are few peaks with radius less than $R_f$, this suggests
that the probability distribution of peak sizes is
fairly narrow, most peaks having a radius around the average.

We would like to compare the average peak radius $R_f$ with the average
radius
of a collapsed region. Equivalently, we would like to compare
the filtering volume $V_f$ with the
average volume of a collapsed region.
The latter is is equal to the volume
fraction occupied by the collapsed regions, \eq{gauss}, divided by their number
density $n\sub{coll}$. In general we do not know $n\sub{coll}$, but we
do know the peak number density $n\sub{peak}$, so we can calculated
the average volume
$V\sub{peak}$ \it per peak \rm of
a collapsed region. It is useful to give the result
as a fraction to the filtering volume $V_f$,
\be
V\sub{peak}/V_f
=\frac12\text{erfc} (\nu/\sqrt2)/
(n\sub{peak}V_f) \label{volume}
\ee
The corresponding mass fraction is roughly $M\sub{peak}
/M =(1+\delta_c)V\sub{peak}/V_f$. The
volume fraction is plotted in figure 14, and one sees that
except for large $\nu$ it is bigger than 1.

At large $\nu$, the ratio may be calculated using
the approximation $n\sub{up}=n_\chi/2$ together with \eq{gauss},
\bea
V\sub{peak}/V_f
\simeqa (2\pi)\threehalf (\langle k^2 \rangle/3)\mthreehalf \nu\mthree/V_f \\
\eqa
\rfrac{3}{R_f^2 \langle k^2 \rangle\half}\threehalf \nu\mthree \label{vpea}
\eea
The fact that it is small indicates that a
typical collapsed region is sitting
on top of a peak, as opposed to the situation for smaller $\nu$ where
it encompasses most of the peak. This is natural, because for
large $\nu$ a collapsed region is exceptionally
high.

\subsection{The number density $n(>M)$}

The main application of these results is to estimate the number density
$n(>M)$ of gravitationally bound systems with mass bigger than $M$, at a given
epoch before $z\sub{nl}(M)$. The systems
are supposed to be identifiable by looking at
the linearly evolved density contrast $\delta(M,\bfa x)$. Each collapsed
region, defined as one in which $\delta(M,\bfa x)>\delta_c$, is supposed to
contain one or more systems with mass bigger than $M$. As we have just seen
this cannot be correct for large $\nu$, but
let us accept it for the moment.

If each collapsed region is identified with a single system, then
$n(>M)=n\sub{coll}$. In general this recipe is useless for lack of an
expression for $n\sub{coll}$. A different prescription, which does lead to a
calculable expression, is to identify each peak within a collapsed region with
a different collapsed object,
\be n(>M)= n\sub{peak} \simeq n\sub{up} \label{peak} \ee
This estimate (usually without the simplifying second equality) is
widely used in the literature.
It is certainly
the same as the estimate $n(>M)=n\sub{coll}$ for large $\nu$, where we know
that there is just one peak per collapsed region. To what extent the
prescriptions are the same for lower $\nu$ is not known, because the number of
peaks per collapsed region is not known.

If at some epoch the linearly evolved density contrast does have many
peaks within a collapsed region, an interesting
situation arises. At a somewhat earlier epoch, $\delta(M,\bfa x)$ was smaller,
and a separate contour $\delta(M,\bfa x)=\delta_c$ was wrapped around each
peak.
In other words, each peak of the linearly evolved density contrast,
filtered on scale $M$, was inside a single collapsed region, and presumably
represented a separate gravitationally bound system. At the later epoch when
the collapsed region encompasses many peaks of the linearly evolved density
contrast, we have a bigger gravitationally bound system. If the original
systems
survive, the identification of each peak with a separate system is correct,
but it misses the larger system which contains the original systems. Of
course, missing this one system does not affect the total count much, so if
this case is typical of collapsed regions containing many peaks the estimate
$n(>M)=n\sub{peak}$ is better than the estimate $n(>M)=n\sub{coll}$. If, on
the other hand, the original systems have merged, that identification is
wrong, and the whole of the collapsed region should be identified with just
one
gravitationally bound system. If this case is typical, the estimate $n(>M)
=n\sub{coll}$ would be better, if only we had a formula for it. Which
case is the more likely? A clue is provided by the observation made earlier,
that if there are several peaks in a collapsed region they typically seem to
lie near the surface of the region, a part of the surface wrapping itself
around each peak. This picture would suggest that the estimate
$n(>M)=n\sub{peak}$ is the more reasonable, the peaks of the linearly evolved
density contrast in a typical collapsed region representing structures which
have not existed long enough to merge.

A different prescription was used by Press and Schechter (1974), to derive a
widely used alternative formula. They worked with the differential number
density,
\be \frac{{\rm d}n}{{\rm d}M}\equiv \frac{\rm d}{{\rm d}M} n(>M)
\label{dndm} \ee
At a given epoch, if the filtering mass $M$ is increased by an amount ${\rm
d}M$ then $\nu$ is increased by an amount ${\rm d}\nu$, and the volume
fraction occupied by the collapsed regions is reduced by an amount ${\rm d}V$
given by \eq{dgauss}. Press and Schechter suppose that the eliminated volume
consists of objects with mass between $M$ and $M+{\rm d}M$, corresponding to
the idealisation that filtering the density contrast on any mass scale $M$
cuts out precisely those objects with mass less than $M$ while leaving
unaffected objects with mass bigger than $M$. Ignoring the overdensity
$\simeq(1+\delta_c)$ of the collapsed regions this implies that the number
density ${\rm d}n$ of such objects is given by
\bea M\frac{{\rm d}n}{{\rm d}M}\eqa \left[M\frac{{\rm d}(R_f^2)}{{\rm d}M}
	\right] \frac{{\rm d}(\sigma^2(M))}{{\rm d}(R_f^2)} \frac{{\rm d}\nu}
	{{\rm d}(\sigma^2(M))} \frac{{\rm d}V}{{\rm d}\nu} \frac{{\rm d}n}
	{{\rm d}V} \\
	\eqa \left[\frac{2R_f^2}{3} \right]
\left[-\sigma^2(M)\langle k^2 \rangle\right]
	\left[-\frac{\nu}{2\sigma^2(M)} \right]
	\left[\frac1{\sqrt{2\pi}}e^{-\nu^2/2}\right]
	\left[\frac1{V_f}\right]\\
	\eqa \frac{R_f^2 \langle k^2
\rangle}{3}\frac{1}{4\pi^2 R_f^3} \nu e^{-\nu^2/2}
	\label{psch} \eea

Press and Schechter multiplied this formula by a factor 2, so that when
integrated over all masses it would give the  total mass density of the
universe, rather than just the half corresponding to the regions of space
where the linearly evolved density contrast is positive. They thus arrived at
the estimate
\be n(>M)\simeq n\sub{ps}
\equiv\int_M^\infty
\frac{\langle k^2 \rangle\pr}{6\pi^2 R_f\pr}
\nu\pr e^{-\nu^{\prime 2}/2} \frac{dM\pr}{M\pr} \label{nps} \ee
In this equation, $R_f\pr=R_f(M\pr)$, and similarly for
$<k^2>\pr$ and $\nu\pr$.
The factor 2 inserted by Press and Schechter is not justified by their
argument, because the linearly evolved density contrast has nothing to do with
reality in the non-linear regime $\sigma(M)>1$. On the other hand, the
neglected overdensity gives a factor $\simeq(1+\delta_c)=2$ to 3. Thus the
factor 2 goes in the right direction, and the Press-Schechter formula is
reasonably well founded theoretically.

{}From \eq{nufirst},
\be \frac{\nu\pr}{\nu}=\frac{\sigma(M)}{\sigma(M\pr)} \ee
The right hand side is independent of
$b$, $\delta_c$ and $z$, so it
follows that $n\sub{ps}$, like $n\sub{peak}$, depends on these
quantities through $\nu $, which involves the combination
$b\delta_c(1+z)$. In figure 15 is plotted the ratio of the
two alternative estimates
$n(>M)=n\sub{ps}$ and $n(>M)=n\sub{peak}$,
for $M=10^{10}\msun$ and for $M=10^{15}\msun$.
One sees that the estimates agree to better than a factor 2 for
$\nu\lsim 2$.
Presumably, this indicates that in this regime
the assumptions underlying the two
estimates are compatible, in that increasing $M$ by a small amount
cuts out portions of the collapsed regions which have mass of order $M$
and are centred on peaks with height of order $\nu(M)$.

For large $\nu$, the Press-Schechter estimate falls below
$n\sub{peak}$. This
can be understood analytically, from the expression
\be \frac{{\rm d}n\sub{peak}}{{\rm d}M}
\simeq \frac{{\rm d}n\sub{up}}{{\rm d}M}
=\pd{n\sub{up}}{\nu}
	\frac{{\rm d}\nu}{{\rm d}M} + \pd{n\sub{up}}{\langle k^2 \rangle}
	\frac{{\rm d}\langle k^2 \rangle}{{\rm d}M} + \pd{n\sub{up}}{\gamma}
	\frac{{\rm d}\gamma}{{\rm d}M}
\ee
The first term dominates for large $\nu$, leading to the ratio
\be \frac{{\rm d}n\sub{ps}/{\rm d}M}{{\rm d}n\sub{peak}/{\rm d}M}=2\rfrac{3}
	{R_f^2\langle k^2\rangle} \threehalf\nu\mthree
\label{ratio} \ee
Apart from the factor 2, this is just the
filter volume divided by the average volume of a collapsed region
(\eq{vpea}).

Both estimates are too
big when $\nu$ is big. The estimate $n\sub{peak}(\simeq n\sub{coll}$)
is too big
because it counts many `collapsed regions' which have mass less than the
filtering mass and do not therefore contribute to
$n(>M)$.
To obtain an accurate
estimate, one would have to excise such regions before evaluating
$n\sub{peak}$.
The Press-Schechter
estimate is too big
because it assumes that all of the reduction in volume of the
collapsed regions in going from $M$ to $M+\delta M$ corresponds to structures
with mass in this range, whereas some of the eliminated volume will belong to
collapsed regions with mass less than $M$. To obtain an accurate
estimate, one would again have to excise these regions before
calculating the eliminated volume
and hence obtaining the Press-Schechter estimate.
Neither of these effects can be quantified analytically, because
relevant statistical results are not known. The fact that the ratio
\eq{ratio} is 2 times the ratio of the filtering volume to the
average volume of
a collapsed region presumably indicates that the regions
which ought to be excised dominate both $n\sub{peak}$ and the
reduction in volume leading to the Press-Schechter estimate.
The implication is that even the latter is substantially too high
at large $\nu$.

In what follows we focus the estimate $n(>M)=n\sub{peak}
(\simeq n\sub
{up}$) because of its simplicity, but give some results also for
$n\sub{ps}$.

\subsection{Bias factors}

Before comparing the number densities with observation, we need one more
result from the analysis of BBKS, namely an estimate of the galaxy and galaxy
cluster bias factors at the present epoch. According to the CDM cosmogony such
a bias factor will be present for any class of objects, if they form at an
epoch when
$\nu$ is
significantly bigger than 1 and therefore originate as exceptionally
high peaks of the density contrast.
In general the
bias occurs partly through the
excess clustering of these peaks, and
partly through additional
non-linear clustering after the objects have formed.
An estimate including both effects is
\be b\sub{obj}=1+\frac{\tilde\nu}{\sigma(M)}\label{bbks}\ee
where
\be
\tilde\nu=\nu-\frac{\gamma\theta(\gamma,\gamma \nu)}{1-\gamma^2}
\ee
In these expressions, everything on the right hand side is
to be evaluated at the epoch when the objects form.

\subsection{Comparison with observation}

The prediction $n(>M)=n\sub{peak}$ is plotted against redshift in figure 16,
for masses $M=10^{15}\msun$, $10^{12}\msun$, $10^{10}\msun$ and
$10^8\msun$. For each case,
three curves are given corresponding to the parameter choices
$\{n,b\}=\{1,1\}$, $\{1,1.6\}$ and $\{.7,1.6\}$.
Each curve ends at the epoch $z\sub{nl}(M)$, when the linear approach
ceases to be valid.

Let us consider first the case  $M=10^{15}\msun$, which corresponds to very
large galaxy clusters.  Since this mass corresponds to the normalisation
scale $R_f\sim10h\mone\Mpc$ there is little dependence on $n$ in the
range $.7<n<1$. There is however strong dependence on $b$,
and several authors
(Kaiser 1984; Bardeen {\it et al} 1986; Bardeen {\it et
al} 1987; Dalton {\it et al} 1992; Nichol {\it et al} 1992;
Efstathiou, Bond and White 1992; Adams \it et al \rm 1992)
have claimed that $b\gsim 1.6$ is preferred over
$b=1$. Let us first reproduce this result, then comment on its
uncertainty.

Since $\sigma(10^{15}\msun)=.8/b$, we assume that the
filtered density contrast on this scale
is still evolving linearly at the present epoch.
 The quantities of interest are
the number density  $n(>M)$ and the ratio $b_c/b$ of the
cluster bias factor to the galaxy bias factor.
The prediction for $n(>M)$
is $5.0\times 10^{-6}\Mpc\mthree$ if $b=1$, and $1.0\times
10^{-6}\Mpc\mthree$ if $b=1.6$.
Following for instance BBKS and Bardeen {\it et al} (1987), we
identify clusters in this mass range with Abell clusters of richness class
$>1$, and hence observed number density $7.5\times10^{-7}\Mpc\mthree$ (Bahcall
\& Soniera 1983). Thus, the prediction for $n(>M)$
is about right if $b=1.6$, but too big
if $b=1$. Coming to $b_c/b$,
 one finds from \eq{bbks} that
if $b=1.6$, then $b_c=4.0$ giving $b_c/b=2.5$. If, on the other hand,
$b=1.0$, then $b_c=1.5$ which gives $b_c/b=1.5$.
Recent estimates (Dalton {\it et al} 1992;
Nichol {\it et al} 1992) give $\xi_{cc}(r)=(r_0/r)^{1.9\pm.3}$ with
$r_0=(13\pm5)h\mone\Mpc$. Dividing by the galaxy correlation function
$\xi_{gg}=(5h\mone\Mpc/r)^{1.8}$ gives therefore $b_c/b_g=2.4\pm.9$
as an observational estimate. Again, $b=1.6$ is
preferred.

So far so good, but what about the uncertainty? Unfortunately it is
big. First, suppose that we take $\delta_c$ to be equal to $1.69$,
instead of the $1.33$ used in the above estimates. For $b=1$ (1.6)
this multiplies
$n(>M)$ by a factor .56 (.16), and multiplies
$b_c$ by a factor $1.4$ ($1.4$). Second, suppose instead that we multiply
$M$ by a factor $2$, on the ground that there may be this
amount of uncertainty in the observational value of the masses of galaxy
clusters (and remembering also that filtering the density contrast
on mass scale $M$ does not completely eliminate all structure with mass
less than $M$). This multiplies $n(>M)$ by a factor $.30$ ($.09$) and
multiplies
$b_c$ by a factor $1.6$ ($1.7$). Third, do neither of these things
but replace $n\sub{peak}$ by $n\sub{ps}$ which is lower
(though not low enough remember) for large $\nu$. This multiplies
$n(>M)$ by a factor $.78$ ($.35$) without changing $b_c$.
Each of these not unreasonable changes has a significant
effect
on the predictions. Finally, implement them all simultaneously.
This multiplies $n(>M)$ by a factor $.03$ ($.0004$),
and multiplies $b_c$ by a factor 2.3 (2.4). The effect is to
make $b=1$ strongly preferred over $b=1.6$!

Our conclusion is that one cannot yet reliably constrain
$b$ by considering the present number density and bias factor
of rich clusters. On the other hand, these quantities are certainly very
sensitive to $b$, and would constrain it well if the uncertainties could
be removed.
Similar remarks probably apply to earlier epochs, but as yet the data
are too sparse to say much.

For galaxies, the linear epoch ends before the present. The number density
$n(>M)$ at the end of the non-linear epoch cannot be estimated reliably,
and it will also evolve with time because of
merging and other non-linear phenomena.
As a result, no reliable prediction is possible for the quantity $n(>M)$ at
the present epoch.

However, in the biassed theory of galaxy formation, the formation of {\it
luminous} galaxies is supposed to stop during the linear regime, at some epoch
given roughly by \eq{bbks}. By demanding that this value reproduces
the value of $b$ used to normalise the amplitude one obtains
the epoch of luminous galaxy formation, and hence the observed
number density, as a function of $b$. This approach has been implemented
for $n=1$
by BBKS and by Bardeen \it et al \rm (1987).
Here we extend the calculation to $n=.7$,
and comment on the uncertainty.

The stars in figure 16b
for $b=1.6$ indicate the epoch of formation of luminous galaxies
which is needed to reproduce this bias factor,
according to \eq{bbks}. This epoch is $z=4.7$ (2.8) for
$n=1$ (.7), and the corresponding value of $n(>M)$ is $4.8\times 10
\mfour\Mpc\mthree$ ($7.9\times 10\mfour\Mpc\mthree$).
Ignoring merging etc., this comoving number density
should be
equal to the presently observed number density $n_g(>M)$ of
luminous galaxies for
$M=10^{12}\msun$.
An observational estimate of $n_g(>M)$
is the Schechter parametrisation, which taking the ratio
of luminosity $L$ to mass $M$ to be independent of $M$ is (Ellis {\it et al}
1988)
\be
M \frac{{\rm d}n_g}{{\rm d}M} =\phi_*\rfrac{M}{M_*}^{-.07} e^{-M/M_*}
\ee
In this formula
\be
\phi_*=1.56\times10\mtwo h^3 \Mpc \ee
Integrating $M {\rm d}n_g/{\rm d}M$ over all masses gives $M_*=
1.6\Omega\sub{gal}\times 10^{13} h\mone\msun$,
where $\Omega\sub{gal}$ is the contribution to $\Omega$ of luminous
galaxies (including the dark halos) and we
take
$\Omega\sub{gal}=.1$. For $M=10^{12}\msun$, this gives an observational
estimate $n_g(>M)=1.75\times 10\mthree\Mpc\mthree$, which is marked by an
arrow on the $y$-axis of figure 16a.

According to this calculation, the
predicted number density of luminous galaxies is somewhat too small
for $b=1.6$, and in fact one needs a value $b\simeq 1.2$ to $1.3$ to
reproduce it. But now set $M=
5\times 10^{11}\msun$ in the theoretical calculation, on the ground
that the observational value of $M$ could be a factor 2 too high.
With $b=1.6$ this
gives a somewhat earlier epoch of formation $z=5.9$ (3.6)
for $n=1$ (.7).
The corresponding number densities are $n_g(>M)=5.8
\times 10\mfour$ ($1.2\times 10\mthree$), which look much more healthy,
and the actual value of $b$ needed for consistency is now around $1.4$.
Even without looking at any other sources of uncertainty
(such as galaxy merging, which BBKS emphasise),
it seems clear that it is very difficult to pin down
$b$ within the range $1<b<1.6$ from these considerations.

One conclusion, though, does seem fairly robust, which is that
going from $n=1$ to $n=.7$ makes the redshift of bright galaxy
formation
later by around 2 units. As evidence accumulates from $N$-body
simulations and from observation, this difference may eventually
be able to pin down the required value of $n$.

Finally,
consider the results for $M=10^{10}\msun$
and $10^8\msun$, as shown in figures 16c and d.
The observed number densities $n_g(>M)$
according to the Schechter
parametrisation are again shown by arrows (though they become
increasingly uncertain as the mass is reduced).
Within the biassed galaxy formation theory one expects luminous galaxies
with these masses to form
before the epoch $z\sub{nl}(M)$, but the observational bias factor for
them is not known, and there is no reason why it should be equal to
$b$ which refers to bright galaxies. However, the epoch of formation
would need to be a lot earlier than $z\sub{nl}(M)$ to
give agreement with the observed $n_g(>M)$. It is presumably
more reasonable to suppose that merging has taken place,
reducing $n_g(>M)$ from its original value. Again, there
does not seem to be a useful constraint on the parameters $n$ and
$b$.

\subsubsection*{The quasar density}

So far we have focussed on observations at small redshift.
One can also ask about observation
at $z\gsim 1$. In particular, quasars
have now been seen out to a redshift of about 5, and are the oldest
observed objects in the universe. In order to produce the observed luminosity,
some galaxies must have evolved to contain a sufficient concentration of
mass-energy. One then needs to estimate the minimum mass required and the
number density, in order to discover if the CDM cosmogony can explain the
observations.  Such a comparison was made for $n=1$ by Efstathiou
and Rees (1988). Using both the $n\sub{peak}$ and
Press-Schechter prescriptions they calculated
$n(>M)$ for various masses as a function of redshift.
Our results are far higher than theirs, but the difference can
be explained if they chose
$b=2.5$ and defined $b$ with respect to
$J_3$ normalisation instead of $\sigma_g$ normalisation
(they are not explicit about these choices, but our calculation
essentially agrees with theirs if these changes are made in it).
As in the case of large clusters at the present epoch,
the extreme sensitivity to normalisation is caused by the
fact that $\nu$ is substantially bigger than 1.

The number density of quasars assumed by Efstathiou and Rees was $10^{-6}
h^{3}$ Mpc$^{-3}$, out to a redshift of around $4$. Since then, observations
have become more stringent, and to be consistent with present data one
requires this number density out to a redshift of $5$ (Martin Rees, private
communication). This value also assumes that the quasar lifetimes are not too
short, in which case one needs multiple generations. One also needs to know
the mass which is required to be evolving nonlinearly to harbour the quasar,
and they estimated at the time that $10^{12} \msun$ was the smallest that one
could safely get away with. However, recent work (Martin Rees, private
communication) has suggested that only $10^{10} \msun$ is required, which
corresponds to a large loosening of the constraint.
As one sees from figures 16b or 16c, the
upshot of all this is that enough quasars may form
at high redshift, even if $n=.7$ and $b=1.6$.

Another cosmological requirement on high redshifts is the Gunn-Peterson
constraint on the amount of neutral hydrogen in the inter-galactic medium. In
the CDM cosmogony this implies that some structure has formed before $z=
5$,
in order to re-ionise the hydrogen (Schneider, Schmidt \& Gunn 1989), but it
is not clear how much or of what kind. Even with $n=.7$, structure with
$M\lsim 10^{10}\msun$ forms at the epoch $1+z\simeq 9/b$ so a value for $b$ as
high as $1.6$ need not be problematical.

\subsection{The virial velocity in a galaxy or cluster}

The linear approach has been pushed further by some authors, to try
to predict the virial velocity $v(M)$ of structures of mass
$M$
(Blumenthal {\it
et al} 1984; Bardeen {\it et al} 1987; Evrard 1989; Henry \& Arnaud 1991;
Evrard \& Henry 1991; Adams et al 1992). The virial velocity
of a gravitationally bound system
is defined as the rms velocity of its
constituents in the centre of mass frame. According to the virial
theorem it is
given by \be v^2=\frac{GM}{R_g} \label{virial}
\ee where $M$ the mass of the system
and $R_g$ is its
gravitational radius, defined by the requirement that its potential energy is
$-GM^2/R_g$.
The idea is to relate
$R_g$ to the comoving size $R\sub{com}$ of the object in the early
universe, defined by
\be M=(4\pi/3) R\sub{com}^3 \rho_0 \label{mass} \ee
where $\rho_0=3H_0^2/(8\pi G )$ is the present mass density.
To obtain such a relation one can use the spherical collapse model of
Section 6.2, in the following manner.
First,
assume that the system has virialised, with no energy loss,
before the epoch $z\sub{form}$ when in the absence of pressure forces the
system   would  have contracted to a point. Next, assume that the system
collapsed from an initial configuration at rest, with a density profile of the
same shape as the profile after virialisation and without energy
loss (dissipation). It then follows from the virial
theorem, \eq{virial}, that $R_g$ is equal to
one half of the initial gravitational
radius. Finally, set the initial gravitational radius equal to the initial
radius of the edge of the object, defined as the sphere containing
mass $M$. Using the results already quoted in section
\ref{NONLIN}.2  this gives
\bea R_g\mthree\eqa 8\frac{9\pi^2}{16} R\sub{com}\mthree
	(1+z\sub{max})\mthree \\
\eqa 32\frac{9\pi^2}{16} R\sub{com}\mthree (1+z\sub{form})\mthree\\
	\eqa 178 R\sub{com}\mthree  (1+z\sub{form})\mthree \eea
Then \eqs{virial}{mass} give
\be \rfrac{v}{126\km\,\s\mone}^2=\rfrac{M}{10^{12}\msun}\twothird
(1+z\sub{form}) \label{sthtw} \ee
This is the expression quoted by several of the  authors
mentioned above (Evrard 1989; Henry \& Arnaud
1991; Evrard \& Henry 1991), and the others presumably used a similar
expression though they are less explicit.
Obviously, the assumptions leading to this relation
are at best extremely crude approximations. It seems clear,
in fact, that \it any \rm unique relation between the mass, virial velocity
and formation epoch of gravitationally bound systems can only be a rough
approximation.

In figure 16, the arrows on the horizontal axes show $z\sub{form}(v)$,
calculated from  \eq{sthtw}, corresponding to estimates of the upper and
lower observational limits on observational values of $v$ for objects
of mass $M$.
These estimates of $v$ are extremely crude. For galaxies, they correspond
roughly to those given by Blumenthal {\it et al} (1984) without any attempt to
update that analysis in the light of more recent observations. For clusters,
the upper limit corresponds roughly to X-ray observations (Henry \& Arnaud
1991); the observational lower limit of about $1000\km\,\s\mone$ would in that
case correspond to negative $z$, a value of about $1300\km\,\s\mone$
corresponding to $z=0$.

The most direct way of utilising the relation $z\sub{form}(M,v)$
is to set $z\sub{form}$ equal to $z\sub{nl}(M)$. This gives
the virial velocity $v(M)$ of systems
formed when the scale $M$ goes nonlinear.
Blumenthal \it
et al \rm (1984) identified such systems with luminous galaxies and
with galaxy clusters,
but according to the biassed galaxy formation
theory that identification is
wrong in the former case. Alternatively, the relation can be combined
with a theoretical
estimate of the number density $dn(M,z)$ of objects with mass
between $M$ and $M+dM$, which form at epochs between \rm $z$ and
$z+dz$, to calculate the
number density
$n(z,>v)$ of objects with virial velocity bigger than
$v$, which exist at redshift $z$.
This is essentially the approach taken by the other authors
mentioned above.

We saw in Section 6.7 that even an estimate of $n(z,>M)$ is subject
to large uncertainties, with the result that
it cannot be used to reliably constrain
the parameters
$n$ and $b$ within the range
allowed by the QDOT and COBE data.
An estimate of $n(z,>v)$ inherits
uncertainties of the same
type, plus additional ones coming from the use of
the relation $z\sub{form}(M,v)$. It
is unclear how to quantifying the total
uncertainty, but it seems likely that it will again prevent
one from constraining $n$ and $b$ within the allowed range.

\subsection{Summary}

In this long section we have discussed galaxies and clusters, making the
maximum use of linear theory despite the fact that these objects are
gravitationally bound. Our discussion has extended earlier $n=1$
calculations to the case $.7<n<1$. It has treated some points of
principle not considered in earlier discussions, and in some cases
has made for the first time a crude estimate  of the uncertainty in
the calculations.
Despite, or perhaps because of,  these
improvements the conclusion is that they do not provide
significant additional constraints on the pair of parameters $n$ and $b$,
beyond those already provided by the
QDOT and COBE data.

\section{Discussion and Conclusions}
\label{CON}
\setcounter{equation}{0}
\def\theequation{\thesection.\arabic{equation}}

As small scale structure does not seem to provide any additional constraints
over the large scale results of sections \ref{MWB} and \ref{CLUSTER}, our
final constraints are those illustrated in figure 11. From this, we conclude
that for natural inflation one must have $n > 0.70$ (which is a separate
$2$-sigma exclusion on two pieces of data). If one is willing to push all the
observations to their limits and fold in extra uncertainties such as allowing
$h < 0.5$, it is just conceivable that such a model allows a fit to the
clustering data such as APM. The most important requirement for this is that
the microwave fluctuation level should be at the top of its permitted range.
One sees from the spread of the dashed lines on figure 11 that if the COBE
result is exactly correct then the limit on $n$ will immediately be pushed
well above 0.8.

For power-law and extended inflation, the limit is already $n > 0.84$ at high
confidence. This clearly excludes the possibility that these models can fit
the clustering data, and indeed we have already remarked (Liddle \& Lyth 1992)
that in fact this limit rules out simple forms of extended inflation, as they
give a big bubble constraint (Liddle \& Wands 1991) $n < 0.75$. (Changing from
a cold dark matter cosmogony to a different choice also does not seem to
salvage the situation (Liddle \& Lyth 1992).) The strength of this constraint
is due in the main to the large contribution of gravitational waves to the
microwave anisotropies at low $n$; $n=0.84$ is in fact coincidentally close to
the break-even point where gravitational and scalar modes contribute the same
amount. Note once again that the limit will strengthen dramatically if the
microwave result comes down. If the COBE result is exact, then that will push
$n$ well above 0.9.

Assuming one takes the QDOT result seriously, then the bias is also strongly
constrained with a $2$-sigma maximum of around $1.6$. For both types of
inflationary model, the lowest allowed $n$ has this maximum bias. Hence the
most optimistic power-law model, which may just fit the clustering data,
appears to be a natural inflation style model with values of $n \sim 0.70$ and
$b \sim 1.6$. In addition to an improvement on clustering issues, such a model
also seems to deal adequately with the formation of structure (contrary to our
more pessimistic assessment in LLS).

A preference has been expressed for models with $n$ in the range $0.8$ to
$0.9$ (see {\it eg} Salopek 1992; Davis {\it et al} 1992; Lucchin, Matarrese
\& Mollerach 1992), because these models are claimed as allowing a higher bias
than the $n=1$ model. However, with the current error range on the COBE data
our figure 11 shows little benefit in this at present, because even at $n=1$
the COBE error bars allow quite substantial biasses at $n=1$ (and remember
this is without including that the spectrum from say a chaotic inflation model
has a slight but not insignificant slope and also mildly significant
gravitational wave contributions). Even at $n=1$ it appears that QDOT rather
than COBE provides the upper limit to the bias. However, as soon as the upper
limit to the fluctuations comes down, either by tightened error bars or
independent measurement, this effect may become more relevant.

To conclude, we have found strong constraints on the slope of the primeval
spectrum when generated by various inflationary models. Most extended
inflation models appear to be ruled out completely. Power-law inflation is
viable, but only for values of $n \gsim 0.84$, too high to allow an
explanation of the clustering data. Natural inflation (and related) models are
the most promising candidates for generating useful power-law spectra ---
provided the true level of fluctuations is close to the top of the COBE range
they seem marginally able to explain the excess large scale clustering as now
seen in many optical surveys. In most aspects, such a model does at least as
well as a bias one standard CDM model, with the advantage of a more plausible
cluster abundance as well as helping with galaxy clustering statistics. We
must note however that even without including corrections to the flat spectrum
and a gravitational wave contribution, even standard CDM appears compatible
with biasses up to $1.4$ or $1.5$. Such a model may however have difficulties
with the clustering data.

\vspace{12pt}
{\em Final Note:} As we were completing this paper, we received preprints by
Cen {\it et al} (1992) which discusses various astrophysical consequences of
power-law spectrum including $N$-body results (but without considering
gravitational waves), and by Adams {\it et al} (1992) discussing the
astrophysics with particular emphasis on the natural inflation scenario. Where
we cover the same ground as these papers, our results appear to be in good
agreement.

%%%%%%%%%%%%%%%%%%%%%%%%%%%%%%%%%%%%%%%%%%%%%%%%%%%%%%%%%%%%%%%%%%%%%%
\section*{Acknowledgements}

We would like to specially thank Will Sutherland for the collaboration
reported in LLS (some of the results of which are reproduced here). We also
wish to thank Peter Coles, Ed Copeland, Rod Davies, Richard Frewin, Josh
Frieman, Martin Hendry, Rocky Kolb, Tony Lasenby, Jim Lidsey, Lauro
Moscardini, Martin Rees, David Salopek, David Spergel, Paul Steinhardt, Andy
Taylor, Peter Thomas and Simon White for helpful discussions and comments at
various stages of this work. ARL is supported by the SERC, and acknowledges
the use of the STARLINK computer system at the University of Sussex.
%%%%%%%%%%%%%%%%%%%%%%%%%%%%%%%%%%%%%%%%%%%%%%%%%%%%%%%%%%%%%%%%%%%%%%
\section*{References}
\frenchspacing
\begin{description}
\item Abbott, L. F. and Wise, M. B. 1984a, {\it Astrophys J Lett} {\bf 282},
	L47.
\item Abbott, L. F. and Wise, M. B. 1984b, {\it Nucl Phys} {\bf B244}, 541.
\item Adams, F. C., Bond, J. R., Freese, K., Frieman, J. A. and
	Olinto, A. V. 1992, ``Natural Inflation: Particle Physics Models,
	Power Law Spectra for Large Scale Structure and Constraints from
	COBE'', CITA preprint.
\item Albrecht, A. and Stebbins, A. 1992, ``Cosmic String with a Light
	Massive Neutrino'', Fermilab preprint.
\item Babul, A. and White, S. D. M. 1991, {\it Mon Not Roy astr Soc} {\bf
	253}, 31p.
\item Bahcall, N. and Soniera, R., 1983, {\it Astrophys J} {\bf 270}, 20.
\item Bardeen, J. M., Bond, J. R., Kaiser, N. and Szalay, A. S. 1986, {\it
	Astrophys J} {\bf 304}, 15 [BBKS].
\item Bardeen, J. M., Bond J. R. and Efstathiou, G. 1987, {\it Astrophys
	 J} {\bf 321}, 28.
\item Barrow, J. D. 1987, {\it Phys Lett} {\bf B187}, 12.
\item Bertschinger, E. 1985, {\it Astrophys J Supp} {\bf 58}, 39.
\item Bertschinger, E. and Dekel, A. 1989, {\it Astrophys J Lett} {\bf 336},
	L5.
\item Bertschinger, E., Dekel, A., Faber, S. M., Dressler, A. and
	Burstein, D. 1990, {\it Astrophys J} {\bf 364}, 370.
\item Bertschinger, E. and Gelb, J. M. 1991, {\it Computers Phys} {\bf 5}, 164.
\item Blumenthal, G. R., Faber, S. M., Primack, J. R. and Rees, M. J. 1984,
	{\it Nature} {\bf 311}, 517.
\item Bond, J. R. and Efstathiou, G. 1984, {\it Astrophys J} {\bf 285}, L45.
\item Bond, J. R. and Efstathiou, G. 1987, {\it Mon Not Roy astr Soc} {\bf
	226}, 655.
\item Bond, J. R. and Efstathiou, G. 1991, {\it Phys Lett} {\bf B265}, 245.
\item Bower, R. G., Coles, P., Frenk, C. S., White, S. D. M. 1992,
	``Cooperative Galaxy Formation and Large-Scale Structure'', Durham
	preprint.
\item Brainerd, T. G. and Villumsen, J. V. 1992, \it Astrophys J
\bf 394, \rm 409.
\item Carlberg, R. G. and Couchman, H. M. P. 1989, {\it Astrophys J} {\bf
	340}, 47.
\item Carlberg, R. G., Couchman, H. M. P. and Thomas, P. A. 1990, {\it
	Astrophys J} {\bf  352}, L29.
\item Cen, R., Gnedin, N. Y., Kofman, L. A. and Ostriker, J. P. 1992, ``A
	Tilted Cold Dark Matter Cosmological Scenario'', Princeton preprint.
\item Couchman, H. M. P. and Carlberg, R. G. 1992, {\it Astrophys J} {\bf
	389}, 453.
\item Dalton, G. B., Efstathiou, G., Maddox, S. J. and Sutherland, W. J.
	1992, {\it Astrophys J Lett} {\bf 390}, L1.
\item Davis, M. and Peebles, P. J. E. 1983, {\it Astrophys J} {\bf 267}, 465.
\item Davis, M., Efstathiou, G., Frenk, C. S. and White, S. D. M. 1985, {\it
	Astrophys J} {\bf 292}, 371.
\item Davis, M., Efstathiou, G., Frenk, C. S. \& White, S. D. M.
1992, \it Nature \bf 356, \rm 489.
\item Davis, M., Summers, F. J. and Schlegel, D. 1992 ``Large Scale Structure
	with Mixed Dark Matter'', Berkeley preprint.
\item Davis, R. L., Hodges, H. M., Smoot, G. F., Steinhardt, P. J. and
	Turner, M. S. 1992 ``Cosmic Microwave Background Probes Models of
	Inflation'', Fermilab preprint FERMILAB-PUB-92/168-A.
\item Dekel, A., Bertschinger, E. and Faber, S. M. 1990, {\it Astrophys J}
	{\bf 364}, 349.
\item Dekel, A. 1991, in ``Observational Tests of Cosmological
	Inflation'', eds Shanks, T. {\it et al}, Kluwer academic,
        and in Annals of the New York Academy of Sciences, Volume
        647 (1991).
\item Ellis, R. S. and Peterson, B. A. 1988, {\it Mon
	Not Roy astr Soc } {\bf 232}, 431.
\item Efstathiou, G. and Rees, M. J. 1988, {\it Mon Not Roy astr Soc} {\bf
	230}, 5p.
\item Efstathiou, G. 1990, in ``The Physics of the Early Universe'', eds
	Heavens, A., Peacock, J. and Davies, A., SUSSP publications.
\item Efstathiou, G., Sutherland, W. J. and Maddox, S. J. 1990, {\it Nature}
	{\bf 348}, 705.
\item Efstathiou, G. 1991, in ``Observational Tests of Cosmological
	Inflation'', eds Shanks, T. {\it et al}, Kluwer academic.
\item Efstathiou, G., Bond, J. R. and White, S. D. M. 1992, ``COBE Background
	Radiation Anisotropies and Large Scale Structure in the Universe'',
	Oxford preprint.
\item Evrard, A. 1989, {\it Astrophys J } {\bf 341}, L71.
\item Evrard, A. and Henry, J. P. 1991, {\it Astrophys J} {\bfa 383},
95.
\item Fabbri, R., Lucchin, F. and Matarrese, S. 1986, {\it Phys Lett} {\bf
	B166} 49.
\item Freese, K., Frieman, J. A. and Olinto, A. V. 1990, {\it Phys Rev Lett}
	{\bf 65}, 3233.
\item Frenk, C. S., White, S. D. M., Davis, M. and Efstathiou, G. 1988, {\it
	Astrophys J} {\bf 327}, 507.
\item Fukugita, M., Futamase, T., Kasai, M. and Turner, E. L. 1992, {\it
	Astrophys J} {\bf 393}, 3.
\item Gott, J. R., Melott, A. L. \& Dickinson, M. 1986, {\it Astrophys J}
	{\bf 306}, 341.
\item Gunn, J. E. and Gott, J. R. 1972, {\it Astrophys J} {\bf 176}, 1.
\item Guth, A. H. and Jain, B. 1992, {\it Phys Rev D}{\bf 45}, 426.
\item Henry, J. P. and Arnaud, K. A. 1991, {\it Astrophys J} {\bf 372},
	410.
\item Hodges, H. M. and Blumenthal, G. R. 1990, {\it Phys Rev D}{42}, 3329.
\item Holtzmann, J. 1989, {\it Astrophys J Supp} {\bf 71}, 1.
\item Hogan, C. J., Kaiser, N. and Rees, M. J. 1982, {\it Phil Trans R Soc
	Lond} {\bf A307}, 97.
\item Kaiser, N. 1984, {\it Astrophys J Lett} {\bf 284}, L9.
\item Kashlinsky, A. and Jones, B. J. T. 1991, {\it Nature} {\bf 349}, 753.
\item Kashlinsky, A. 1992, {\it Astrophys J Lett} {\bf 387}, L5.
\item Kolb, E. W. and Turner, M. S. 1990, \sl The Early Universe
	\rm (Addison-Wesley).
\item Kolb, E. W., Salopek, D. S. and Turner, M. S. 1990, {\it Phys Rev D}{\bf
	42}, 3925.
\item Kolb, E. W. 1991, {\it Physica Scripta} {\bf T36}, 199.
\item Krauss, L. M. and White, M. 1992 \it Phys Rev Lett \bf 69,
\rm 869 (1992).
\item La, D. and Steinhardt, P. J. 1989, {\it Phys Rev Lett} {\bf 62}, 376.
\item Liddle, A. R. 1989, {\it Phys Lett} {\bf B220}, 502.
\item Liddle, A. R. and Wands, D. 1991, {\it Mon Not Roy astr Soc} {\bf 253},
	637.
\item Liddle, A. R., Lyth, D. H. and Sutherland, W. J. 1992, {\it Phys Lett}
	{\bf B279}, 244 [LLS].
\item Liddle, A. R. and Lyth, D. H. 1992, ``COBE, Gravitational Waves,
	Inflation and Extended Inflation'', to appear, {\it Phys Lett B}.
\item Lidsey, J. E. and Coles, P. 1992, ``Inflation, Gravitational Waves and
	the Cosmic Microwave Background: Reconciling CDM With COBE?'', to
	appear, {\it Mon Not Roy astr Soc}.
\item Lilje, P. B. 1992, {\it Astrophys J Lett} {\bf 386}, L33.
\item Linde, A. 1983, {\it Phys Lett} {\bf B129}, 177.
\item Linde, A. 1987, in ``300 Years of Gravitation'', eds Hawking, S. W. and
	Israel, W. I., Cambridge University Press.
\item Linde, A. D. 1990, \sl Particle Physics and Cosmology \rm
	(Gordon and Breach).
\item Lucchin, F. and Matarrese, S. 1985, {\it Phys Rev D}{\bf 32}, 1316.
\item Lucchin, F., Matarrese, S. and Mollerach, S. 1992, ``The
	Gravitational-Wave Contribution to CMB Anisotropies and the
	Amplitude of Mass Fluctuations from COBE Results'', Fermilab preprint
	FERMILAB-PUB-92/185-A.
\item Lyth, D. H. and Stewart, E. D. 1990, {\it Astrophys J} {\bf 361}, 343.
\item Lyth, D. H. and Stewart, E. D. 1992a, {\it Phys Lett} {\bf B274}, 168.
\item Lyth, D. H. and Stewart, E. D. 1992b, in preparation.
\item Maddox, S. J., Efstathiou, G., Sutherland, W. J. and Loveday, J. 1990,
	{\it Mon Not Roy astr Soc} {\bf 242}, 43p.
\item Maddox, S. J., Sutherland, W. J., Efstathiou, G., Loveday, J.
	and Peterson, B. A. 1991, {\it Mon Not Roy astr Soc} {\bf 247}, 1p.
\item Mathewson, D. S., Ford, V. L. and Buchhorn M. 1992, {\it Astrophys J
	Lett} {\bf 389}, L5.
\item Melott, A. C. 1990, {\it Physics Reports} {\bf 193}, 1.
\item Mukhanov, V. F., Feldman, H. A. and Brandenberger, R. H. 1992, {\it
	Physics Reports} {\bf 215}, 203.
\item Nichol, R. C., Collins, C. A., Guzzo, L. and Lumsden, S. L. 1992
	{\it Mon Not R astr Soc} {\bf 255}, 21p.
\item Park, C., Gott, J. R. and da Costa, L. N. 1992, {\it Astrophys J Lett}
	{\bf 392}, L51.
\item Peacock, J. A. and Heavens, A. F. 1990, {\it Mon Not Roy astr Soc} {\bf
	243}, 133.
\item Peebles, P. J. E. 1970, \it Astrophys J \bf 75, \rm 13.
\item Peebles, P. J. E. 1980, \sl The Large Scale Structure of the Universe
	\rm (Princeton University Press).
\item Peebles, P. J. E. 1982, {\it Astrophys J Lett} {\bf 263}, L1.
\item Press, W. H. and Schechter, P. 1974, {\it Astrophys J} {\bf 187},
	452.
\item Sachs, R. K. and Wolfe, A. M. 1967, {\it Astrophys J} {\bf 147}, 73.
\item Salopek, D. S., Bond, J. R. and Bardeen, J. M. 1989, {\it Phys Rev
	D}{\bf 40} 1753.
\item Salopek, D. S. 1992, ``Consequences of the COBE Satellite for the
	Inflationary Scenario'', Cambridge preprint.
\item Saunders, W. {\it et al} 1991, {\it Nature} {\bf 349}, 32.
\item Saunders, W, Rowan-Robinson, M and Lawrence, A. 1992, ``The Spatial
	Correlation Function of IRAS Galaxies on Small and Intermediate
	Scales'', QMW preprint.
\item Scaramella, R. and Vittorio, N. 1988, {\it Astrophys J Lett} {\bf 331}
	L53.
\item Scaramella, R. and Vittorio, N. 1990, {\it Astrophys J} {\bf 353}, 372.
\item Scaramella, R. 1992, {\it Astrophys J Lett} {\bf 390}, L57.
\item Schaefer, R. K. and Shafi, Q. 1992, ``Inflation and Large Scale
	Structure Formation after COBE'', ICTP preprint IC/92/118.
\item Schneider, D. P., Schmidt, M. and Gunn, J. E., 1989
{\it Astron J} \bf 98, \rm 1507.
\item Smoot, G. F. {\it et al} 1992, ``Structure in the COBE DMR First Year
	Maps'', NASA preprint.
\item Souradeep, T. and Sahni, V., 1992  ``Density Perturbations,
Gravity Waves and the Cosmic Microwave Background'',
IUCAA preprint.
\item Suto, Y. and Fujita, M. 1990, {\it Astrophys J} {\bf 360}, 7.
\item Suto, Y., Gouda, N. and Sugiyama, N. 1990, {\it Astrophys J Supp}
	{\bf 74}, 665.
\item Suto, Y., Cen, R. and Ostriker, J. P. 1992 \it Astrophys
J \bf 395, \rm 1.
\item Taylor, A. N. and Rowan-Robinson, M. 1992, ``Cosmological Constraints
	on the Spectrum of Density Fluctuations from COBE and QDOT'', QMW
	preprint.
\item Tormen, G., Lucchin, F. and Matarrese, S. 1992, {\it Astrophys J}
	{\bf 386}, 1.
\item Tormen, G. Moscardini, L., Lucchin, F. and Matarrese, S., 1992,
	``The Galaxy Velocity Field and CDM Models'', in preparation.
\item Vittorio, N., Matarrese, S. and Lucchin, F. 1988, {\it Astrophys J}
	{\bf 328}, 69.
\item Vogeley, M. S., Park, C., Geller, M. J. and Huchra, J. P. 1992, {\it
	Astrophys J Lett} {\bf 391}, L5.
\item White, S. D. M., Frenk, C. S. and Davis, M. 1983, {\it Astrophys J Lett}
	{\bf 274}, L1; {\it Astrophys J} {\bf 287}, 1.
\item White, S. D. M., Frenk, C. S., Davis, M. and Efstathiou, G. 1987, {\it
	Astrophys J} {\bf 313}, 505.
\item Wright, E. L. {\it et al} 1992, ``Interpretation of the CMB Anisotropy
	Detected by the COBE DMR'', NASA preprint.
\end{description}
%%%%%%%%%%%%%%%%%%%%%%%%%%%%%%%%%%%%%%%%%%%%%%%%%%%%%%%%%%%%%%%%%%%%%%
\nonfrenchspacing
\newpage
\section*{Figure Captions}

{\em Figure 1}\\
A selection of transfer functions, from Efstathiou (1990) (used in this
paper), Bardeen {\it et al} (1986), Davis {\it et al} (1985) and Holtzmann
(1989), given as appropriate to CDM universes with $\Omega = 1$, $h = 0.5$,
low baryonic content and three neutrino species. They feature $4$, $5$, $3$
and $4$ fitting parameters respectively.

\vspace{0.3cm}
\noindent
{\em Figure 2}\\
The normalisation $b\delta_H(k=1 \Mpc\mone)$ as a function of $n$. The solid
line is from the $\sigma_8$ normalisation (used in this paper), while the
dashed line shows the rival $J_3$ normalisation.

\vspace{0.3cm}
\noindent
{\em Figure 3}\\
The present day spectra, as calculated in the linear approximation, for a
selection of values of $n$. One sees the additional large scale power and the
deficit on short scales when one compares $n < 1$ to the standard CDM
spectrum. Note $k$ is in Mpc$^{-1}$, without a factor of $h$.

\vspace{0.3cm}
\noindent
{\em Figure 4}\\
The dispersion $b\sigma(M)$ as a function of mass, for $n=1$ and $n=0.7$, and
with both choices of filter. The top hat filtered spectra are unity at $M \sim
10^{15}\msun$, as required by the normalisation. The gaussian filtering gives
significantly smaller answers than does the top hat, as its smearing gives a
higher contribution to the larger scales.

\vspace{0.3cm}
\noindent
{\em Figure 5}\\
The triangles indicate the prediction for the mean quadrupole as a function of
$n$ for power-law inflation, and the stars for natural inflation. The dotted
lines plot the COBE observation for two choices of bias (note the bias
implicit in the $y$-axis). The CDM prediction for the mean at bias 1 is very
close to the COBE result. The vertical bars on the starred data indicate the
spread of the {\em pdf} for the quadrupole (for clarity the bars have been
omitted for the triangles --- they are the same size); 95\% of the {\em pdf}
is above the bottom of the bars, while 95\% is below the top. The $\chi_5^2$
distribution is not symmetric, so the bars are skewed to higher values
(somewhat concealed by the log plot). A value of $n$ is allowed at 95\%
exclusion if the observations cut through its vertical bar. Modelling the
observational errors (see text) gives an even looser criterion.

\vspace{0.3cm}
\noindent
{\em Figure 6}\\
The predictions for $b \left. \frac{\Delta T}{T}\right|_{10^0}$, as defined in
text, along with the COBE limits (at $1$-sigma) for bias $1$ and $2$. The
upper line represents the power-law inflation predictions, the lower those for
natural inflation. As for the quadrupole, the $n=1$ prediction at bias $1$ is
very close to the COBE result.

\newpage
\noindent
{\em Figure 7}\\
The predicted angular correlation functions for a choice of $n$ are plotted
alongside the observational data from the APM survey (Maddox {\it et al}
1990). With the anticipated residual systematics, values of $n$ between about
0.3 and 0.6 provide reasonable fits, while the standard CDM curve falls well
below the data.

\vspace{0.3cm}
\noindent
{\em Figure 8}\\
A comparison of models with different $n$ against a standard CDM model but
with $\Omega h = 0.2$.

\vspace{0.3cm}
\noindent
{\em Figure 9}\\
The predicted {\em rms} velocity flows, when smoothed with a top hat of radius
$R_f$, for different choices of $n$.

\vspace{0.3cm}
\noindent
{\em Figure 10}\\
The predicted {\em rms} velocity flows in a configuration mimicking the POTENT
observational data. The velocity field is first smoothed with a $12h^{-1}$ Mpc
gaussian, reducing the short scale power, and then smoothed with top hat
filters of radius $R_f$, giving predictions smaller than in figure 9. The
solid lines indicate the predictions for $n=1$ and $n=0.7$. The stars indicate
the POTENT observational data at bias $1$ (read from figure 4 of Dekel
(1991)), and the triangles the same at bias $1.6$. The error bars on the data
(the last ones just overlap) are $1$-sigma. Finally, we emphasise that the
theoretical curves are averages over all observer points, whereas the
observations are a single realisation, with correlated errors due to long
wavelength domination of bulk flows.

\vspace{0.3cm}
\noindent
{\em Figure 11}\\
The limits on the bias $b$ as a function of $n$. The dotted line shows an
upper limit on $n$ to fit the APM data. All other lines are $2$-sigma limits.
The dot-dashed lines indicate the range allowed by QDOT, and apply to all
models. The solid and dashed lines indicate the limits from COBE for power-law
(and extended) inflation and natural inflation respectively. For power-law
inflation, one has $n > 0.84$ and $b < 1.5$, which rules out extended
inflation which needs $n < 0.75$ to satisfy the bubbles constraint. For
natural inflation, one has $n > 0.70$ and $b < 1.6$. This is well outside our
APM fit range (though it is close to the range allowed by Efstathiou, Bond \&
White (1992)). If one does not attempt to fit optical galaxy clustering, then
the allowed region is that marked with the star, bounded by the microwave
limits appropriate to a given scenario.

\vspace{0.3cm}
\noindent
{\em Figure 12}\\
The asphericity parameter $x\mone$ plotted against $\nu$.

\vspace{0.3cm}
\noindent
{\em Figure 13}\\
The ratio $2n\sub{up}/n_\chi$ plotted against $\nu$.

\vspace{0.3cm}
\noindent
{\em Figure 14}\\
The ratio $V\sub{peak}/V_f$ plotted against $\nu$.

\vspace{0.3cm}
\noindent
{\em Figure 15}\\
The ratio $n\sub{up}/n\sub{P-S}$ plotted versus $\nu$.
The ratio is not very sensitive to $n$, and only the case
$n=1$ is plotted.

\vspace{0.3cm}
\noindent
{\em Figure 16}\\
The number density $n(>M)$ as a function of redshift. It is calculated
theoretically by equating it with the number density of peaks of the linearly
evolved density contrast, filtered on the mass scale $M$. Figures 16a--d
refer respectively to $M=10^{15}\msun$, $10^{12}\msun$ $10^{10}\msun$ and
$10^8\msun$. For each case there are three curves. They correspond to the
three choices $\{n,b\}=\{.7,1.6\},\{1,1.6\}$ and $\{1,1\}$ which mark the
corners of the more or less triangular region allowed by the QDOT and COBE
data (figure 11). The arrows on the vertical axes give the observed galaxy
number densities at the present epoch. The arrows on the horizontal axes
indicate very roughly the `observed' range of formation epochs, deduced from
the indicated range of virial velocities. Each curve ends at the epoch when
$\sigma(M)=1$, signalling the end of linear evolution. According to the theory
of biassed galaxy formation, luminous galaxies form significantly before that
epoch. In the case of bright galaxies, the epoch can be calculated by
demanding that the resulting bias factor is equal to $b$, and it is indicated
by a star in figure 16 b. The theoretical and observational uncertainties in
are discussed in the text.

\end{document}